\documentclass[12pt,preprint]{elsarticle}

\usepackage{amsmath,amssymb,amsfonts, bbm}
\usepackage{amsthm}
\usepackage{graphicx}
\usepackage{geometry}
\usepackage{placeins}
	\usepackage{xargs}                      %
	\usepackage[pdftex,dvipsnames]{xcolor}  %
\usepackage{caption}
\usepackage{subcaption}
\usepackage{algorithm}
\usepackage{algpseudocode}

	\usepackage{physics} %

\newproof{pf}{Proof}
\newproof{pot}{Proof of Theorem \ref{thm2}}

\usepackage{hyperref}%
\usepackage{cleveref}

 \usepackage{xurl}%
\hypersetup{
    colorlinks=true,
    linkcolor=blue,
    filecolor=magenta,      
    urlcolor=cyan,
    }
    
\newcommand{\inR}{\in\mathbb{R}^}
\usepackage{dsfont}

\newcommand{\ncp}{\kappa_{NN}} %
\usepackage{enumitem}
\usepackage[scr=boondoxo]{mathalfa}

\begin{document}
	\begin{frontmatter}

    \title{%
    Closed-loop training of static output feedback neural network controllers for large systems: A distillation case study}
\author[inst1]{Evren Mert Turan}

\affiliation[inst1]{organization={Department of Chemical Engineering, Norwegian University of Science and Technology (NTNU)},%
            addressline={Sem Sælandsvei 4}, 
            city={Trondheim},
            postcode={7491}, 
            country={Norway}}

\author[inst1]{Johannes Jäschke}

	\begin{abstract}
The online implementation of model predictive control for constrained multivariate systems has two main disadvantages: it requires an estimate of the entire model state and an optimisation problem must be solved online. These issues have typically been treated separately. This work proposes an integrated approach for the offline training of an output feedback neural network controller in closed loop. Online this neural network controller computers the plant inputs cheaply using noisy measurements. In addition, the controller can be trained to only make use of certain predefined measurements. Further, a heuristic approach is proposed to perform the automatic selection of important measurements. The proposed method is demonstrated by extensive simulations using a non-linear distillation column model of 50 states. 

\end{abstract}

\begin{keyword}
Optimal control \sep Neural networks \sep Closed-loop optimal controllers \sep output feedback controller\sep measurement noise\sep Explicit MPC\sep End-to-end learning
\end{keyword}

\end{frontmatter}

\section{Introduction}

Due to the difficulty of explicitly defining a control law for multivariate interacting systems with constraints, a popular approach is to use model predictive control (MPC) to implicitly define a feedback control law. Given an estimate of a system's state, MPC uses a dynamic model to forecast the behaviour of the system subject to a sequence of control actions, optimises to find this sequence to minimise an objective and satisfy system constraints. This sequence of control actions results from an open-loop optimisation, i.e. the third control action does not take into account the sequence of disturbances that may have occurred. It is well established that use of open loop optimisation can have dramatic effects due to stochasticity or innacuracies in the system forecast. MPC incorporates feedback by only implementing the first control action, before repeating the optimisation with a new estimate of the system state. 

Thus, MPC incorporates feedback due to measurements that are used to estimate the system state. However, each time a control action is determined, MPC requires (a) the current state of the system to be estimated from measurements, and (b) the solution of an optimisation problem. For many systems one can use a linear(-ised) model which allows for computing the state estimate and optimal solution reliably and efficiently. Some systems are inherently non-linear and the use of linearised models can result in poor controller performance. However, non-linear models generally incur a significantly larger computational cost, and as such there may be a non-negligible delay between measurement and control action. This is especially true for large systems, or if uncertainty is considered. This delay reduces performance and can (if significant enough) destabilise the system \citep{santos2001line, findeisen2004computational}.

Various approaches have been proposed to reduce the computational cost associated with MPC. 
This can be done by general approaches, such as improving the optimisation algorithm, or more tailored approaches such as using the time between measurements to compute the control sequence in advance, and then adjust this based on the new measurement \citep{diehl2005real,zavala2009advanced,jaschke2014fast}.
Many authors have taken motivation from recent advances in machine learning to reduce the computational cost of MPC. The main idea is to approximate some control relevant mapping that is typically unavailable or expensive to compute. Examples of these mapping include: terms in the objective function \citep{seel2022convex,Turan2023-convex-terminal}, steady state inverses of processes \citep{ramchandran1995_nn_inverse}, %
and offline control policies \citep{Parisini1995, Karg2020, Kumar2021, bonzanini2021fast, Drgona2022, Turan2023-nn-policy-jpc}. This work focuses on the latter, where the aim is to learn an explicit output feedback control policy offline instead of solving an optimal control problem online.  

For a ``standard'' linear MPC problem the optimal control law is piecewise affine on polyhedrals, and can be computed offline by solving a parametric program \citep{Bemporad2002}. This has become commonly known as "explicit MPC". Unfortunately, the online computational requirements of this method grows exponentially in the number of states and length of control horizon, and hence this approach is not suitable for large systems. Thus, a more generally applicable approach is to train an approximator to describe the control policy. Typically neural networks are used as they are flexible and cheap to evaluate -- e.g. one can find a neural network that is exactly equivalent to the explicit linear MPC solution, while avoiding the exponential scaling of the online evaluation \citep{Karg2020}.

When training a neural network control policy one may either (a) separate the training from the control problem  by first collecting a database of states and control actions, and then train the controller on this database (optimise-then-learn, or imitation learning)  \citep{Parisini1995, Karg2020, Kumar2021},  (b) combine the two problems by embedding the neural network in the control problem and train the controller on the control objective (optimise-and-learn) \citep{Turan2023-nn-policy-jpc, Drgona2022}, or (c) use a reinforcement learning approach. As we wish to find a static, offline policy we do not consider reinforcement learning approaches in this work.

Although the imitation learning approach is conceptually and implementationally simpler, it is restrictive as it requires data for the controller to imitate. In addition, as the control problem is solved separately to the training, a controller that does well on the training/validation data may have poor closed-loop performance \citep{Turan2023-nn-policy-jpc}.  In the optimise-and-learn approach
the controller is directly trained based on the controller performance, and hence avoids problems occured by separating the control and training problems \citep{Turan2023-nn-policy-jpc,li2022}. In addition, if the controller is directly embedded into the system dynamics, e.g. \citet{Turan2022dycops,Turan2023-nn-policy-jpc}, then the training occurs in \textit{closed-loop}. Closed-loop training allows for flexible handling of uncertainty and also allows the use of a smaller controller compared to the open-loop alternative, e.g. \citep{Drgona2022}. In general, the optimise-and-learn approach is a more complicated training problem, however it allows for a more flexible specification of the resulting controller.

In this work we find an explicit output-feedback neural network control policy, that can be cheaply evaluated, using noisy measurements, to give control actions. This policy is trained in a closed-loop  optimise-and-learn formulation \citep{Turan2023-nn-policy-jpc}, which allows considering measurement noise, and pragmatic usage of a subset of only some measurements. We show that one can perform measurement selection as part of the training of the control policy, and highlight some potential issues of this joint approach.  The control policy is demonstrated on simulations of a non-linear distillation column. This is a considerably larger problem (50 states) than all other examples in the optimise-and-learn literature.

Prior literature has predominantly considered relatively small scale examples, with the notable exception of \citet{Kumar2021} which considered an imitation learning approach applied to large linear MPC problem for control of distillation column. 

To the authors knowledge the only work considering offline optimisation of an output feedback neural network  policy is \citep{Drgona2022}, which considers a small single-input single-output system. Unlike this prior work, we incorporate knowledge of the measurement noise in the closed loop training of the policy and consider control policies that do not use all available measurements. In addition, we consider a much larger nonlinear model of a distillation column consisting of 50 states. A key challenge of the training is the size of the state space. Any practical implementation requires restriction of attention to a smaller region of the state space. 
We sample the state space to find the typical operationally relevant region \citep{Kumar2021,bonzanini2021fast}, and train the policy to start from that region. Unlike an imitation which would requires sampling along trajectories starting in this region \citep{bonzanini2021fast}, the control policy is optimised in closed loop and hence these samples are ``implicitly generated'' during training \citep{Turan2023-nn-policy-jpc}.  

This work is structured as follows: we briefly summarise the background and relevant literature in Section \ref{distil-sec: background}, and proceed to formulate the output-feedback training problem in Section \ref{distil-sec: output-feedback}. In Section \ref{distil-sec: distillation-formulation} we introduce the distillation problem. In Section \ref{distil-sec: control-opt-train}  we construct the control policies, and present extensive numerical results of their closed-loop performance in Section \ref{distil-sec: results}.  Lastly in Section \ref{distil-sec: discussion} we discuss important aspects of the results and potential future work. 

\section{Background}\label{distil-sec: background}
\subsection{Model predictive control}
In a continuous time formulation of model predictive control (MPC) we wish to solve the time dynamic optimisation problem\footnote{Note that in practice, most MPC problem are given in a discrete time formulation, but we focus on the continuous time formulation as we are interested in obtaining a continuous time feedback control law.}:
  \begin{subequations}\label{distl-eqn: mpc}
    \begin{align}
           u^*_{MPC}(t,z_0)=\arg\min_{u(t)}\:\:&  J(z_0,u)\\
            &J(z_0,u)= \int_{t_0}^{t_f} l(z,u,t)\:dt + V_f\left(z(t_f),u(t_f)\right)\\
           	&0=f\left(\dot{z}(t),z(t),u(t),p\right)\\
            &z(t) \in \mathcal{Z}\label{distl-eqn: state-con}\\
            &u(t) \in \mathcal{U}\\
            &z(t_0) = z_0
    \end{align}
    \end{subequations}
    where $t$ is time, $t_0$ is the initial time, $t_f$ the final time, $z\in \mathcal{Z}\inR{n_z}$ is the system state, with time derivative $\dot{z}$, $u\in \mathcal{U}\inR{n_u}$ is the control input, $u^*_{MPC}$ is the optimal control input, $f$ is an implicit differential equation, $z_0$ is the state at initial time, $l$ is the stage cost, $V_f$ is the terminal cost, and $\mathcal{Z}$ and $\mathcal{U}$ are constraint sets. In process control problems these constraints sets are typically defined by upper and lower bounds constraints on $z$ and $u$. %

    To incorporate feedback MPC is implemented in a receding-horizon approach. Given a state estimate, $z_0$,  \eqref{distl-eqn: mpc} is solved, and the control $u^*_{MPC}(t_0,z_0)$ is implemented. Then at $t_0+\Delta t$, given a new estimate the problem is resolved to find a new control action. Thus, MPC implicitly defines the control policy:
    \begin{equation}
        \kappa_{MPC}(z_0) =  u^*_{MPC}(t_0,z_0)
    \end{equation}
    where $\kappa_{MPC}$ returns the first control action of the open-loop solution. We emphasise that to solve \eqref{distl-eqn: mpc} the full initial state of the system, $z_0$, needs to be specified. In a typical process many of the process variables are unmeasured, and hence have to be estimated from the available noisy measurements.

\subsection{Neural network and control policies}

   When using MPC, due to the need to perform state estimation, and then solve an optimisation problem \eqref{distl-eqn: mpc} there will be a delay between receiving the system measurement and sending the optimal input to the plant. If large enough this delay can reduce the controller performance and can even destabilise the system. Although there are efficient optimisation algorithms for MPC, managing the computational delay can be challenging for large systems, especially when non-linear dynamics and uncertainty is considered. We consider the use of neural network control policies to eliminate this computational delay and provide fast online evaluations for the control actions. 
   
Authors dating back to 1995 \citep{Parisini1995} have proposed the use of neural networks to learn control policies that otherwise require an expensive evaluation such as MPC, as they are universal approximators, i.e. sufficiently large feed-forward neural networks are able to approximate bounded, continuous functions defined on a compact subset of $\mathbb{R}^{n_z}$ to an arbitrarily low tolerance \cite{hornik1989multilayer, cybenko1989approximation}. More recently authors have shown that neural networks with a specific architecture can exactly represent the control policies of linear MPC \citep{Karg2020}, and under some assumptions can approximate nonlinear MPC policies to arbitrary precision \citep{Turan2023-nn-policy-jpc}.

Consider the feed-forward neural network, $\ncp$:
  \begin{subequations}\label{distl-eqn: nn}
    \begin{align}
            &\ncp(z,\theta)= \zeta^{(N_L)}, \qquad \zeta^{0} = z\\
            & \zeta^{(i+1)}= \alpha^{(i)}(W^{(i)}\zeta^{(i)} + b^{(i)}),\qquad i=0,\dots, N_L-1 \\
            & \theta = \texttt{vec}(W^{(0)},\ b^{(0)},\ \dots,\ W^{(N_L-1)},\ b^{(N_L-1)}) 
    \end{align}
    \end{subequations}
   where $N_L$ is the number of layers, $\zeta^{(i)}\inR{w^{(i)}}$ is the latent state of layer $i$, $\alpha^{(i)}$ is an activation function, and $W^{(i)}$ and $b^{(i)}$ are weights and biases that are collected in the vector $\theta$. In this work $\ncp$ is trained in closed-loop to yield a feedback control policy.
    
    \subsubsection{Optimise-and-learn formulation}
    In this paper, we use an optimise-and-learn approach to train the control policy in closed loop \citep{Turan2023-nn-policy-jpc,Turan2022dycops, Drgona2022,li2022}. In this approach the neural network is embedded into the system dynamics to form a single dynamic optimisation problem. This problem is not computationally feasible to solve online, however one can find an offline policy to be used for controlling the system starting in some region $\mathcal{Z}_0$ by solving:
    \begin{subequations}\label{distl-eqn:nn-no-unc-policy}
	\begin{align}
		\min_{\theta}\:\: &\mathop{\mathbb{E}}_{\pi_{z_{0}}} \left[ J(z(t_0),u) \right] \label{distl-eqn: E-obj}	\\
   1 &=\mathop{\mathbb{P}}_{\pi_{z_{0}}} \left[z(t)\in\mathcal{Z}\right]  \label{distl-eqn: prob-constraint}\\
		0&=f(\dot{z}(t), z(t),u(t),p)\\
        u(t) & = \ncp(z(t),\theta), \qquad u(t) \in\mathcal{U}\label{distl-eqn: cntrl-con}\\
   z(t_0) &\sim \pi_{{z}_0}
	\end{align}
     \end{subequations}
     where $\pi_{z_0}$ is some non-zero probability distribution defined on $\mathcal{Z}_0$,  $\mathop{\mathbb{E}}_{\pi_{z_{0}}}$ is the expectation with respect to $\pi_{z_{0}}$, and $\mathop{\mathbb{P}}_{\pi_{z_{0}}}$ is the probability with respect to $\pi_{z_{0}}$. If the control policy defined by the related MPC problem is continuous in $z$ then under mild conditions, minimisation of \eqref{distl-eqn:nn-no-unc-policy} yields the a policy equivalent in performance to the MPC policy \citep{Turan2023-nn-policy-jpc}.
    Typically $u$ is constrained between upper and lower bounds, and thus \eqref{distl-eqn: cntrl-con} can be enforced through the activation function of the final layer. As this is the standard case we assume that the network architecture is chosen to satisfy the constraint. 

    The expectation of the objective and probability constraint need to be approximated in some way to yield a tractable problem.  At each iteration we consider a  stochastic approximation of \eqref{distl-eqn:nn-no-unc-policy}  by evaluating:
    \begin{subequations}\label{distl-eqn:nn-policy-sampled}
	\begin{align}
		\phi&= \sum_{s=1}^{n_s} \omega_sJ(z_s(t_0),u_s(t)) + \rho(z_s(t_0),u_s(t))  	\\
		0&=f(\dot{z_s}(t), z_s(t),u_s(t),p)\\
         u_s(t) & = \ncp(z_s(t),\theta), \qquad u_s(t) \in\mathcal{U}\\
        z_s(t_0) &= \texttt{sample}(\pi_{{z}_0}),\qquad s=1,\dots, n_s
	\end{align}
     \end{subequations}
     where $n_s$ is the number of samples, $z_s(t)$ is the trajectory of sample $s$ taken from $\pi_{{z}_0}$,  $ \omega_s>0$ weights the different trajectories, and $\rho$ is some penalty function used to enforce the constraint \eqref{distl-eqn: state-con} \citep{Turan2023-nn-policy-jpc}.  Note that the $s^{th}$ contribution of $\nabla_\theta \phi$ can be calculated by a standard single shooting approach, i.e. $J(z_s(t_0),u_s(t))$ and $\rho(z_s(t_0),u_s(t))$ can be evaluated by solving a differential equation with the current $\theta$, and the gradient with respect to $\theta$ can then be found by algorithmic differentiation. As each contribution to $\nabla_\theta \phi$  is independent, this computation can be done in parallel, greatly reducing the computational cost of taking multiple samples.

     The key aspect that makes \eqref{distl-eqn:nn-policy-sampled} appealing is that if random samples can be taken such that $\phi$ is an unbiased estimator of the penalised objective of \eqref{distl-eqn:nn-no-unc-policy}, then $\nabla_\theta \phi$ is a stochastic approximation of the gradient of \eqref{distl-eqn:nn-no-unc-policy}. Thus, $\nabla_\theta \phi$  can be used in a stochastic optimization algorithm, e.g. stochastic gradient descent, to solve \eqref{distl-eqn:nn-no-unc-policy}.  There are various options on how to select the samples, weights and number of samples. At one extreme a single sample can be randomly taken at each iteration, and at the other extreme the number of samples, weight, and sample locations can be adapted to maintain a certain approximation tolerance of the expectation and probabilistic constraint (a form of adaptive mini-batching). Although the latter gives a better quality estimation of the gradient at each iteration, each iteration of the former is likely to be much more computationally efficient. 
     
   Note that in some cases the constraint  \eqref{distl-eqn: state-con}  may be enforceable implicitly  by the system dynamics, network architecture or similar. If this is not the case then a penalty approach must be used and adequate satisfaction of the constraint should checked after training. 

     \subsubsection{Selection of initial conditions}

     An important choice in the optimisation-based design of off-line control policies is the region $\mathcal{Z}_0$ on which the policy is developed for. For small systems one may consider selecting $\mathcal{Z}_0=\mathcal{Z}$ , however this becomes computationally impractical for systems with more than a few states. The majority of chemical processes operate in a relatively small region of the total feasible state space \citep{Kumar2021,bonzanini2021fast}. If a model is available then this region can be found by simulating closed-loop trajectories of the controlled system subject to assumed disturbances, e.g. changes in set-points and disturbances in the feed. Then from the simulations one can directly estimate $\mathcal{Z}_0$ and $\pi_{z_0}$. 
\section{Optimisation of an output feedback policy}\label{distil-sec: output-feedback}

The previous literature \citep{Parisini1995,Turan2022dycops,Turan2023-nn-policy-jpc,Kumar2021,Karg2020,bonzanini2021fast} has predominantly  focused on neural network control policies based on the assumption of  noiseless full-state feedback. In general the current state of the system is not available, instead we have measurements $\mathscr{y}\inR{n_\mathscr{y}}$ which are related the state by:
\begin{equation}\label{dist-eqn: noise-meas}
    \mathscr{y}(t)= h(z(t),u(t),p) + \eta(t)
\end{equation}
where $h$ is a (potentially nonlinear) measurement equation, and $\eta\inR{n_\mathscr{y}}$ is a vector of noise and/or biases that influence the measurements. 

In this work, we leverage the optimise-and-learn framework to directly optimise a  control policy that:
\begin{enumerate}[label=(\alph*)]
    \item Incorporates knowledge of the expected measurement noise into the optimisation of the feedback policy.\label{distil: enum-output-feecback-a}
    \item Only uses a subset of the potentially available measurements to control the system.\label{distil: enum-output-feecback-b}
\end{enumerate}
The advantages of \ref{distil: enum-output-feecback-a} is that the control policy will be less sensitive to small perturbations because of noise, which should yield a more robust controller. The advantage of  \ref{distil: enum-output-feecback-b} is more system dependent, but resolves around controlling the desired behaviour of the policy. These two points are developed in the following sections:

\subsection{Noise and uncertainty}

Consider measurements of some of the state subject to additive noise $\eta$ as in \eqref{dist-eqn: noise-meas} with $\eta$ distributed by $\pi_\eta$, some compactly defined multivariate probability distribution. If the noise is significant it should be included in the training of the policy. This can be done by adapting the approach for parametric uncertainty from \citet{Turan2023-nn-policy-jpc}. Consider the  augmented state vector $\bar{z} = [z,\ \eta]$, where $\eta$ has zero dynamics. Let $\eta(t_0)$ be distributed by $\pi_\eta$. During training let the control actions be given by:
\begin{equation}
    u = \ncp( \mathscr{y}(t), \theta)
\end{equation}
As $\eta$ is measurement noise not a stochastic input, it should only influence the controller, i.e. in training the dynamics, objective, etc. should still use $z$.
The objective to be minimised is:
\begin{equation}
    \mathop{\mathbb{E}}_{\pi_{z_{0}}, \pi_\eta} \left[ J(\bar{z}(t_0),u) \right] 	= \mathop{\mathbb{E}}_{\pi_{z_0}} \left[ \mathop{\mathbb{E}}_{\pi_{\eta}} \left[ J(\bar{z}(t_0),u) \right]\right]
\end{equation}
where for  clarity we separate the expectations on the right. Along each trajectory in the training $\eta$ has zero dynamics, i.e. $\eta(t)=\eta(t_0)$, and hence acts as a constant measurement bias. However, the controller cannot learn to ``see'' this constant bias as the controller has no memory, and for any state in $\mathcal{Z}_0$ any $\eta$ can be sampled. 

\subsection{Measurement selection}
In general the entire state of system is rarely measured, and instead typically only a subset of the states are measured. Thus, it is a natural desire to have a controller that only uses this subset of states. In addition, we may also select a subset of these measurements for use. One may interested selecting only some of the available/possible measurements when:
\begin{itemize}
   
    \item Reliable measurements are expensive.
    \item Certain states can only be estimated with high uncertainty.
    \item We wish to influence the behaviour of the control policy, e.g. only specific measurements should influence the output.
    \item We wish to identify which measurements are more important for control.
     \item The system under consideration consists of interacting sections, or is distributed. In this case we may wish that specific control actions are only dependent on a sparse selection of the total measurements.
\end{itemize}

Consider the control policy:
\begin{equation}
    u = \ncp(H\mathscr{y},\theta)
\end{equation}
where $H\inR{n_d\times n_\mathscr{y}}$ is a matrix containing only zero entries and $n_d$ ones on the main diagonal.  By specifying $H$ one specifies which measurements the control policy should use, which can be done using engineering judgement.
Another other approach is to include the selection of measurements (i.e. selecting the non-zero entries of $H$) in the training problem.

\label{distil-sec: input-selection}
One cannot directly optimise for which $n_d$ entries to select as this is a computationally intractable problem. Instead we use elastic net regularisation as an heuristic. For simplicity of presentation we assume that $n_d=n_\mathscr{y}$, i.e. all measurements are available candidates to be used for feedback in the neural network controller.
Let $H$ be a diagonal matrix, and let $\bar{\theta}=\texttt{vec}(H,\ \theta)$. Consider the use of the  regularised objective:
\begin{equation}\label{distil-eqn: elastic-reg}
    \min_{\bar{\theta}}\:\: J(\bar{\theta}) + \lambda_1 (\lambda_2\|\bar{\theta}\|_1 + 0.5(1-\lambda_2)\|\bar{\theta}\|_2^2)
\end{equation}
where $\lambda_1$ controls the regularisation strength, and $0\le\lambda_2\le1$ controls the type of regularisation. $\lambda_2=1$ and $\lambda_2=0$ correspond to $l_1$ (lasso) and $l_2$  (ridge) regularisation respectively.  This regularisation  is called elastic net regularisation, and it promotes small values of $\bar{\theta}$. The $l_1$ term values promotes sparsity in $\bar{\theta}$, while the $l_2$ term penalises large values. Importantly as $H$ is included in $\bar{\theta}$ this regularised objective can be used for input selection as a measurement is unimportant it's corresponding entry in $H$ will be zero or close to zero. Note that the regularisation has to be applied to all of $\bar{\theta}$ as otherwise the network can adjust to use large values of $\theta$ to compensate for the small values in $H$.  After training one can select the important inputs by selecting the entries of $H$ larger than some magnitude. 
Lastly, note that although this heuristic approach can be used to find a policy that uses $n$ measurements, there are no guarantees that the best $n$ measurements will be selected. However, due to the intractability of the exact measurement selection problem this is an attractive heuristic that has been used in similar context \citep{feng2017sparse}. %

\section{Distillation problem formulation}\label{distil-sec: distillation-formulation}
\begin{figure}
    \centering
    \includegraphics[width=0.5\linewidth]{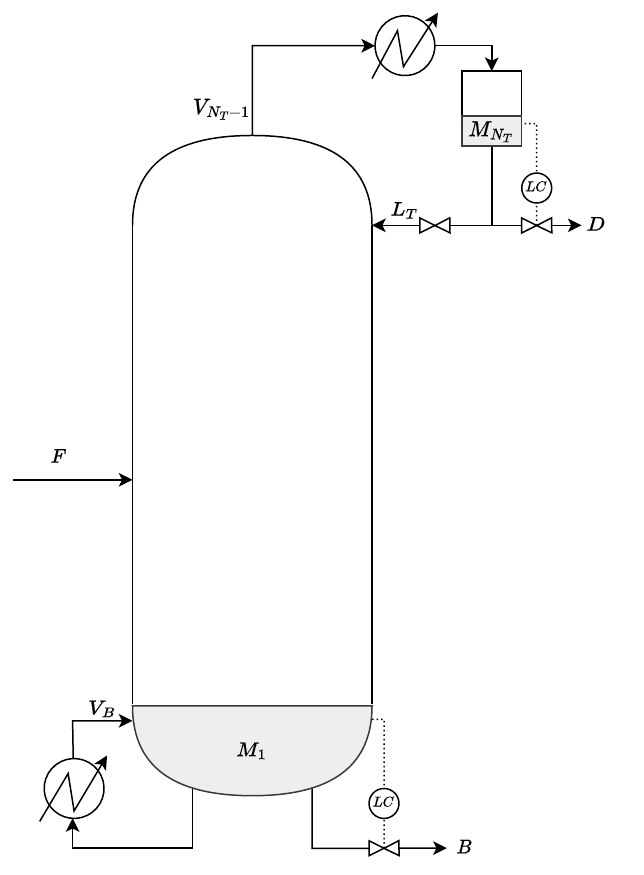}
    \caption{Sketch of a distillation column with LV-configuration and external flows shown. }
    \label{distil-fig:column}
\end{figure}

In this work we consider the separation of an ideal binary mixture by a distillation column of 25 theoretical stages (including reboiler and total condenser, shown in \cref{distil-fig:column}). The desired product composition (for the low boiling component) is 0.99 mole fraction in the top and 0.01 in the bottom. We use the model of \citet{skogestad1997dynamics},  with parameters shown in Table \ref{dist-tab: column-par}. Although the model is relatively simple, it captures the typical dynamic behaviour of a distillation column. For completeness the model is briefly described in Section \ref{distil-subsec: model}. The major assumptions of the model are: constant molar flow rates, constant pressure, constant relative volatility, vapour liquid equilibrium,  negligible vapour holdup, and linearised liquid dynamics. In section \ref{distil-subsec: control-form} aspects of the control problem formulation are described.

Importantly apart from the constraints on the control usage, when integrating the system the inequality and equality constraints of the system are naturally satisfied by the physics of the model. This removes the issue of ensuring constraint satisfaction of \eqref{distl-eqn: state-con}.

\subsection{Distillation model}\label{distil-subsec: model}
The model primarily consists of total and component material balances for each stage of the column. The stages are indexed by $i$, with the bottom stage assigned $i=1$, and the top $i=N_T$.  The balance equations are described first for the trays, then the condenser and last the reboiler. This is followed by the consititutive equations of the column.
\subsubsection{Material balances equations}
 Excluding the feed stage, each tray in the column is described by the  total mole balance:
\begin{equation}\label{dist-eq: total-mb}
    \frac{dM_i}{dt} = L_{i+1} - L_i + V_{i-1} - V_i 
\end{equation}
where $M_i$ is the liquid holdup on stage $i$, $L_i$ is the liquid flow from stage $i$, and $V_i$ is the liquid flow from stage $i$. Additionally, the material balance for the lighter component is:
\begin{subequations}
    \begin{align}
    \frac{dM_ix_i}{dt} &= L_{i+1} x_{i+1} + V_{i-1} y_{i-1} - L_i x_i - V_i y_i \\
         \intertext{which combined with \eqref{dist-eq: total-mb} yields:}
    \frac{dx_i}{dt} &=  \frac{L_{i+1} x_{i+1} + V_{i-1} y_{i-1} - L_i x_i - V_i y_i - x_i (L_{i+1} - L_i + V_{i-1} - V_i)}{M_i} 
\end{align}
\end{subequations}
where $x_i$ and $y_i$ are the mole fraction of the lighter component in the liquid and vapour phases on stage $i$.

Assuming that the feed is mixed directly into feed stage, the total and component mass balances of the feed stage are:
\begin{subequations}
    \begin{align}
        \frac{dM_{N_F}}{dt} =& L_{{N_F}+1} - L_{N_F} + V_{{N_F}-1} - V_{N_F} + F \\
        \frac{dx_{N_F}}{dt}  =& \frac{L_{{N_F}+1} x_{{N_F}+1} + V_{{N_F}-1} y_{{N_F}-1} - L_{N_F} x_{N_F} - V_{N_F} y_{N_F} + Fz_F}{M_{N_F}}\nonumber\\ &- x_{N_F}\frac{L_{{N_F}+1} - L_{N_F} + V_{{N_F}-1} - V_{N_F} + F }{M_{N_F}}
\end{align}
\end{subequations}

 where $F$ is the feed, $N_F$ is the feed stage, and $z_F$ is the mole fraction of the lighter component  in the feed. Similarly, the total and component balances of the condenser are:
 \begin{subequations}
     \begin{align}
        \frac{dM_{N_T}}{dt} &= - L_{N_T} + V_{{N_T}-1} - D \\
        \frac{dx_{N_T}}{dt} & = \frac{V_{N_T} y_{{N_T}-1} - L_{N_T} x_{N_T} - D x_{N_T}}{M_{N_T}}- x_{N_T}\frac{- L_{N_T} + V_{{N_T}-1} - D }{M_{N_T}}
\end{align}
 \end{subequations}
where $D$ is the distillate flow rate. And lastly,  the total and component balances of the reboiler are:
\begin{subequations}
 \begin{align}
        \frac{dM_{1}}{dt} &=  L_{2} - V_1 - B \\
        \frac{dx_{1}}{dt} & = \frac{L_{2} x_{2} - V_1 y_1 - B x_1}{M_{1}}- x_{1}\frac{L_{2} - V_1 - B }{M_{1}}
\end{align}    
 \end{subequations}
where $B$ is the bottoms flow rate.
\subsubsection{Constitutive equations}
By assumption, the liquid flow rates are governed by linearised dynamics:
\begin{subequations}
     \begin{align}
            L_i &= L^0_i + \frac{M_i - M^0_i}{\tau_l} + \lambda(V_{i-1}-V^0_{i-1}),\qquad i=1,\dots,N_{T-1}\\
            L_{NT} &= L_T
     \end{align}
 \end{subequations}

where $L_i^0$ and $M_i^0$  are nominal values of the liquid flow and holdup on stage $i$, $\tau_l$ is the time constant for the liquid flow dynamics and $\lambda$ describes how the vapour flow rate influences the liquid flow rate (the K2-effect \citep{skogestad1997dynamics}). The liquid flow rate from the top tray is simply given by the reflux flow rate, $L_T$.

By the assumption of constant  molar flows and negligible vapour hold-up the vapour flow rates are given by:
\begin{subequations}
     \begin{align}
             V_{i} &= V_{i-1}\qquad  i=2,\dots,N_{T-1},\: i\not = N_F\\
            V_{N_F} &= V_{N_F-1} + (1-q_F)F\\
            V_1 &= V_B
     \end{align}
 \end{subequations}
where $V_B$ is the boilup flow rate, and $q_F$ is the liquid fraction of the feed.

The vapour liquid equilibrium is assumed to given by:
\begin{equation}
    y_i = \frac{\alpha x_i}{1+(\alpha-1)x_i}
\end{equation}
where $\alpha$ is the relative volatility.

To calculate tray temperatures a linear relationship based on the pure component boiling points is assumed:
\begin{equation}
    T_i = x_i T_{b,L} + (1-x_i)T_{b,H}
\end{equation}
where $T_{b,L}$ and $T_{b,H}$ are the boiling points of the light and heavy components.

\begin{table}
\centering
\caption{Summary of nominal column parameters.}
\label{dist-tab: column-par}
\begin{tabular}{lll}

Parameter              &Description& Value   \\ 
\hline \hline
$F$&Feed rate [kmol/min]& 1.0\\
$z_F$&Feed composition& 0.5\\
$q_F$&Feed liquid fraction& 1.0\\ 
 $\alpha$& Relative volatility&1.75\\
 $\tau_L$& Time constant for liquid flow dynamics [min]&0.063\\
 $\lambda$& Constant describing the K2-effect&0.0\\
 $M^0_i$& Nominal liquid holdup[kmol]& 0.5\\
 $L_i^0$& Nominal liquid flow rate[kmol/min]&$\begin{cases}
3.564 i \le N_F\\
2.564 i < N_F
\end{cases}$\\
 $V_i^0$& Nominal vapour flow rate [kmol/min]&3.065\\
 $T_{b,L}$& Light boiling point [K]& 341.9 \\
 $T_{b,H}$& Heavy boiling point [K]& 357.4\\
 
 $K_D$& P-controller tuning for distillate&10.0\\
 $K_B$& P-controller tuning for distillate&10.0\\
 $D^0$& Nominal distillate flow [kmol/min]&0.5\\
 $B^0$& Nominal boilup flow [kmol/min]&0.5\\
 \hline
\end{tabular}
\end{table}

Lastly, we specify that the column is operated in LV-configuration which means that the liquid levels in the reboiler and condenser ($M_1$, $M_{N_T}$) are controlled by the product flows, $B$ and $D$:
\begin{subequations}
    \begin{align}
        D &= D_s + K_D(M_{N_T}-M^0_{N_T})\\
        B &= B_s + K_B(M_{1}-M^0_{1})
    \end{align}
\end{subequations}
The reflux and boilup, $L_T$ and $V_B$, remain as control variables hence the name LV-configuration. This is a reasonable assumption as it is the ``conventional'' choice for distillation columns \citep{skogestad2007_dos_and_donts}. Note that although this controller stabilises the liquid levels, the column itself remains unstable.%

\subsection{Control problem formulation}\label{distil-subsec: control-form}

In the following we describe details of the control problem -- the objective, constraints,  assumed noise of the measurements and MPC parameters.

\subsubsection{Objective and constraints}\label{dist-sec: obj-mpc}
As the objective we consider regulating the product and distillate compositions to their set-points with a small penalty on moving the inputs from the nominal values:
\begin{align}\label{distl-eqn: control-obj}
        J = \int_{t_0}^{t_f} &(x_1(t)-0.01)^2 + (x_{N_T}(t)-0.99)^2 \nonumber\\&+  0.001\left(\left(V_B(t)-V_B^0\right)^2 + \left(L_T(t)-L_T^0\right)^2\right)\: dt
\end{align}
In addition we impose the following inequality constraints:
\begin{subequations}\label{distil-eqn: inequalities}
    \begin{align}
    0&\le x(t) \le 1\\
    0&\le y(t) \le 1\\
    0&\le M(t) \\
    0&\le V_B(t) \le 3.25 \label{distl eqn: vb-con}\\
    0&\le L_T(t) \le 2.75  \label{distl eqn: lt-con}
\end{align}
\end{subequations}
Note that, apart from apart from \cref{distl eqn: vb-con,distl eqn: lt-con}, these constraints are implicitly enforced by the system dynamics, i.e. adaptive time-stepping of a differential equation solver can ensure that the constraints are satisfied. This is similarly the case if the MPC problem is solved with single shooting \citep{Biegler2010}.

\subsubsection{Disturbances, measurement noise and operating region}\label{distl-sec: disturb-and-noise}\label{dist-sec: disturbance sequence}
\begin{figure}
    \centering
    \includegraphics[width=0.8\linewidth]{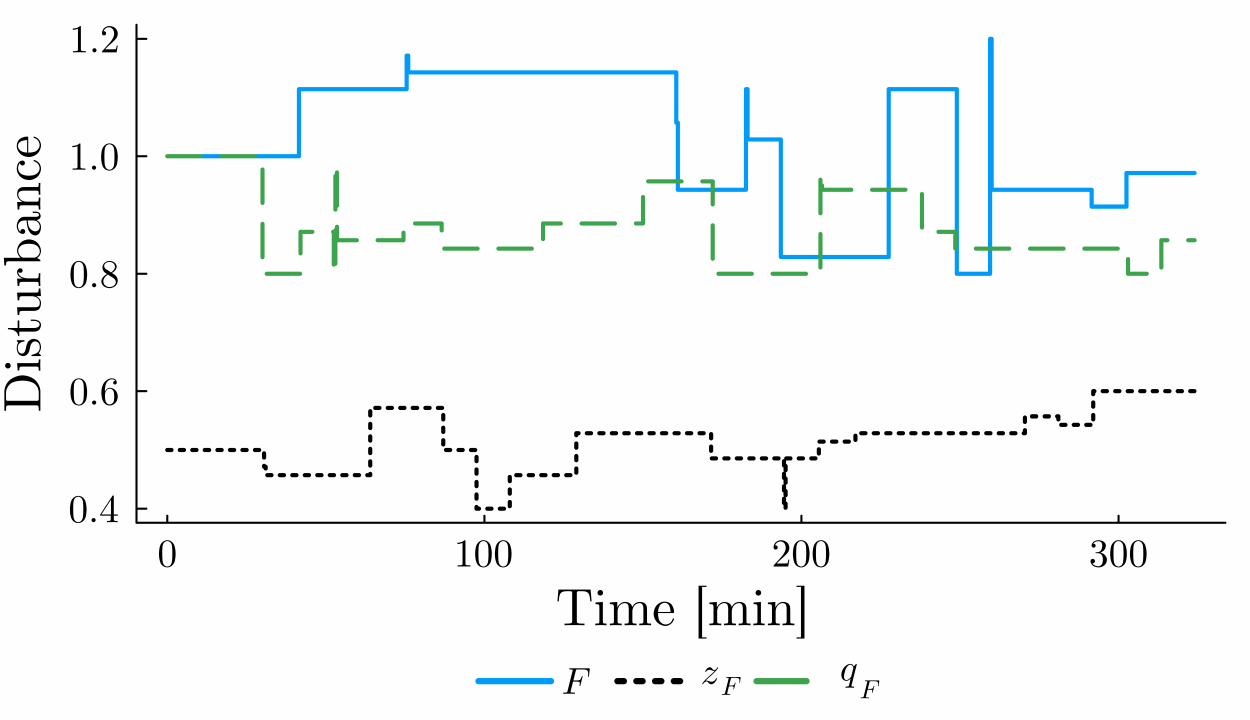}
    \caption{Disturbance profile used in the comparison of the control policies.}
    \label{distil-fig:disturbance-column}
\end{figure}

We assume that potential disturbances, and the ranges, are the column feed flow-rate $[0.8\ 1.2]$, feed composition $[0.4\ 0.6]$ and feed liquid fraction $[0.8\ 1.0]$. Using these disturbance ranges we generate a multi-level pseudo-random sequence of disturbances, and by simulating the distillation column controlled by MPC (see section \ref{distil-sec: benchmark-mpc}) find the effective operating region of the column. 100 disturbances are used with the time between disturbances randomly selected as one of 10 linearly spaced levels between $0.5$ and $10$ minutes, yielding a total simulation time of 485 minutes. At each disturbance time-point one of the disturbances are randomly selected and changed to one of 15 levels equally spaced between the upper and lower bounds of the respective disturbance. The column is initialised with the nominal feed composition  and nominal liquid holdups on each stage and after 15 minutes the disturbance sequence starts. The same approach is used to generate a disturbance sequence to test different controllers against each other, this sequence is shown in Figure \ref{distil-fig:disturbance-column}.  

\begin{figure}
    \centering
    \includegraphics[width=0.8\linewidth]{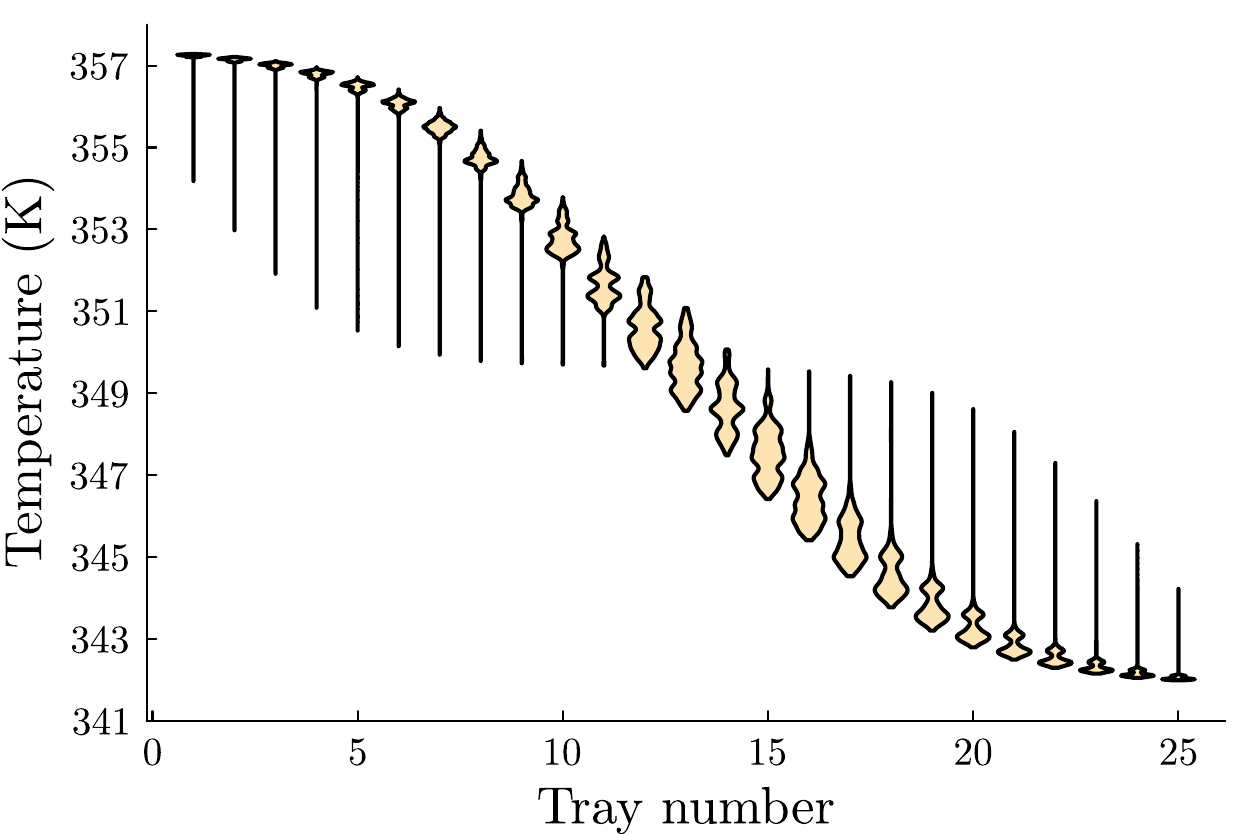}
    \caption{Violin plot showing the temperature range of the distillation column subject to disturbances, to find the relevant operating region of the column.}
    \label{distil-fig:violin-mpc-column}
\end{figure}

The resulting temperature operating range of the column is shown in Figure \ref{distil-fig:violin-mpc-column}. Due to the controller the ends of the column are not significantly influenced by the disturbances while in the middle of the column there is considerable variation. Importantly, from this data (and the corresponding hold-up data) one can see that a very small region of the feasible operating space is visited in standard operation. This reduction of the feasible state space is key aspect of computational feasibility when considering high dimensional systems.

We assume that there are 30 candidate column measurements available: the temperatures on each stage ($T_{1:N_T}$), feed flow rate ($F$), feed temperature ($T_F$), feed liquid fraction  ($q_F$), and the liquid holdups in the reboiler and condenser ($M_1$ and $M_{N_T}$). Because composition measurements are unreliable and typically subject to delays, temperature measurements are normally used instead. All measurements are assumed to be subject to measurement noise as summarised in Table \ref{distil-tab: noise-summary} where $\mathcal{N}(\mu,\sigma)$ denotes the normal distribution with mean and standard deviation $\mu$ and $\sigma$, and the distributions are truncated at the specified min and max values.   %

\begin{table}
\centering
\caption{Summary of assumed measurement noise}
\label{distil-tab: noise-summary}
\begin{tabular}{lll}
\hline
Measurement     & Distribution & (min, max) \\ \hline
Temperature     & $\mathcal{N}(0,\ 0.015)*(T_{bH}-T_{bL})$             &   $(-0.775, 0.775)$         \\
Flow rate       & $\mathcal{N}(0,\ 0.03)$              &    $(-0.1, 0.1)$        \\
Liquid fraction & $\mathcal{N}(0,\ 0.03)$              &     $(-0.1, 0.1)$       \\
Holdup          & $\mathcal{N}(0,\ 0.03)$              &     $(-0.1, 0.1)$       \\ \hline
\end{tabular}
\end{table}
\section{Controller optimisation and training}\label{distil-sec: control-opt-train}
The proposed method is implemented in Julia 1.7, with major use of the following packages: Flux.jl \cite{Flux.jl-2018}, Zygote.jl \cite{innes2018-zygote},   DifferentialEquations.jl  \cite{differentialequations.jl}, and JuMP.jl \cite{JuMP-paper}. For optimisation of the MPC and neural networks Ipopt  \cite{ipopt-Wachter2006} and RMSProp  \cite{RMSProp} are used. 

The control policies we consider are summarised in Table \ref{dist-tab: summary of controllers}. We construct an MPC policy $\kappa_{mpc}$ and four neural network policies. The first two policy are trained using all of the available measurements (note this excludes the tray holdups). We consider training this controller with and without measurement noise included in the training, yielding $\kappa_{all}$ and $\kappa_{all}^{no noise}$ respectively. The second policy,  $\kappa_{reg}$, also has all the available measurements, but is trained with the  elastic net regularisation term \eqref{distil-eqn: elastic-reg}. The training will thus select a reduced set of inputs that the policy will use. The third policy, $\kappa_{sel}$, is trained with only four user specified measurements as inputs using the unregularised  objective. 

The inputs of $\kappa_{sel}$ are chosen as $\zeta = [\bar{T}_5, \bar{T}_{10}, \bar{T}_{16}, \bar{T}_{21}]$.  These temperatures are chosen in emulation of classic control configurations for distillation columns. In such a control scheme instead of controlling the product purities directly, the deviation of a temperature in the bottom and top section of the column from some setpoint is controlled. This is because the temperatures away from the top and bottom are more sensitive to changes in the inputs than the product temperatures \citep{skogestad1997dynamics,skogestad2007_dos_and_donts}. Although we don't directly specify a set-point for these trays, we also do not give $\kappa_{sel}$ the product temperatures. Any control policy that $\kappa_{sel}$ defines must implicitly transform and combine the temperatures and regulate this combination to some set-point. 

After training of the neural network control policies, all the control policies are tested on their performance subject to the disturbance sequence shown in Figure \ref{dist-sec: disturbance sequence}. This comparison is detailed in section \ref{distil-sec: results}.

\begin{table}[]
\centering
\caption{Summary of controllers compared in the distillation case study}
\label{dist-tab: summary of controllers}
\begin{tabular}{lccc}
\hline
      Controller        & Objective                                       & Controller inputs & Training noise \\ \hline
$\kappa_{mpc}$ & Discretised form of \eqref{distl-eqn: control-obj}     & 55 (all states) & N/A \\
$\kappa_{all}^{no\ noise}$  & \eqref{distl-eqn: control-obj}& 30                & No             \\
$\kappa_{all}$         & \eqref{distl-eqn: control-obj}& 30                & Yes            \\
$\kappa_{reg}$ & \eqref{distl-eqn: control-obj} with elastic net penalty & 30              & Yes \\
$\kappa_{sel}$          & \eqref{distl-eqn: control-obj}& 4 temperatures    & Yes            \\ \hline
\end{tabular}
\end{table}

\subsection{The benchmark MPC policy}\label{distil-sec: benchmark-mpc}
\begin{figure}
    \centering
    \includegraphics[width=0.8\linewidth]{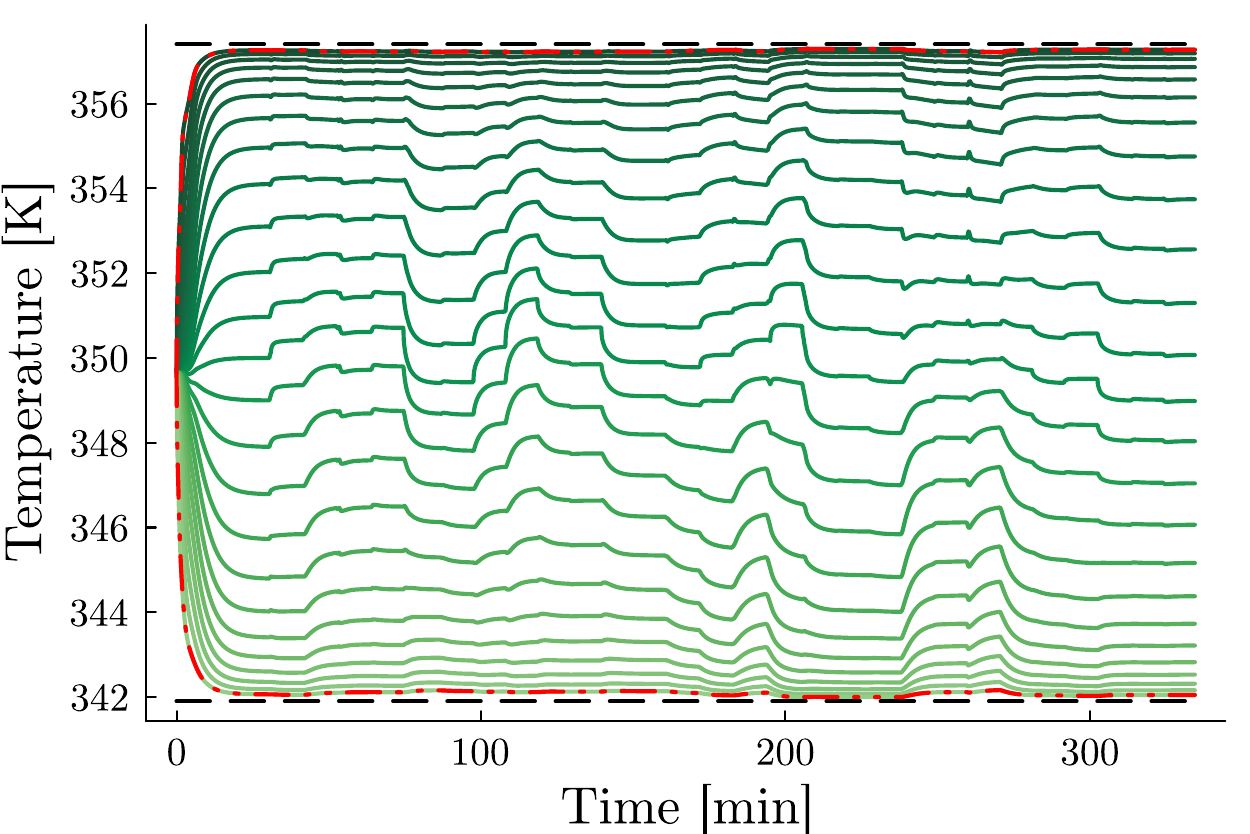}
    \caption{Temperature profiles of distillation column controlled by MPC with test disturbances. The dashed black lines indicate the pure component boiling points, and the red dash-dotted line indicates the temperature in the reboiler and condenser.}
    \label{distil-fig: Tprof-mpc-test}
\end{figure}

We construct a MPC policy to be used a base-line, ``best-case'' control policy. As such we conservatively assume that the MPC has perfect full state feedback. In a more realistic comparison a state estimator would be used with the noisy measurements that are provided to the other control policies. In the MPC problem we use implicit Euler to  discretise the dynamic system. Note that in the discretised problem the objective and constraints are only evaluated at the discrete time points. A discretisation time of 30 seconds, and control horizon of 20 minutes are chosen, thus there are $N_H=41$ points in the time discretisation. The temperature profile of the distillation column when controlled by MPC  is shown in Figure \ref{distil-fig: Tprof-mpc-test}. Note that although the column temperatures are in continuous time, the control output of the MPC is piecewise-constant on 30 second intervals.

\subsection{Training control policies}\label{distil-sec: results-training}

\subsubsection{Specification of neural policies}
 We consider optimisation of a two layer feed-forward neural network control policy of the form:
\begin{subequations} \label{distil-eqn: control-policy}
    \begin{align}
        [L_T,\ V_B] &= \frac{1}{2}[2.75,\ 3.25]\circ( 1 + \ncp(H\zeta,\theta))\label{distil-eqn: output-trans}\\
        \alpha^{(i)}(\zeta) &= (1+e^{-\zeta})^{-1},\qquad i=0,1\\
        \bar{\theta}&=\texttt{vec}(H,\ \theta)
\end{align}
\end{subequations}
where $\circ$ is the element-wise product and $H\inR{n_\zeta \times n_\zeta}$ is a diagonal matrix. As discussed in section \ref{distil-sec: input-selection}, the use of $H$ with a regularised objective promotes a sparse selection of the inputs during training. For consistency, when comparing between the different policies we still use $H$ when training with the un-regularised objective. The sigmoid function is used as the activation function for all layers of the network including the neural network output. This choice of activation function means that the controller inequality constraints \eqref{distil-eqn: inequalities} are  satisfied for all $\bar{\theta}$. 

For $\kappa_{all}$ and $\kappa_{reg}$ the policy takes the input vector:
\begin{subequations}     
\begin{align}
        \zeta & = [\bar{T}_1,\ \dots, \bar{T}_{N_T}, F, \bar{T}_F, q_F, M_{1}, M_{N_T}]\\
        \bar{T}_i &= \frac{T_i - T_{bL}}{T_{bH} - T_{bL}}
\end{align}
\end{subequations}
where $\bar{T}$ is the normalised temperature.  A normalised temperature is used as then all elements of the input vector are the same magnitude. For both  $\kappa_{all}$ and $\kappa_{reg}$ we use a network of width 30, which means that there are $\sim1000$ network parameters. 

The inputs of $\kappa_{sel}$ are chosen as $\zeta = [\bar{T}_5, \bar{T}_{10}, \bar{T}_{16}, \bar{T}_{21}]$, in emulation of classic control configurations for distillation columns. We use the same structure of the network but with only 4 nodes in the input layer. To keep the network a similar size ($\sim1000$  parameters) we chose a width of 150.

We note that any control policy found by $\kappa_{sel}$ is within the scope of $\kappa_{reg}$ which is itself within the scope of $\kappa_{all}$. Thus one may expect that $\kappa_{all}$ should perform the best.
However, this is not entirely to be expected. Although inclusion of extra information can theoretically only be used to improve the control, it is well known that despite neural networks being universal approximators they can have worse performance without judicial selection and manipulation of inputs (``feature engineering''). 

\subsubsection{Embedded training of the control policy}
The control policies are embedded in the dynamical system and trained in the optimise-and-learn formulation \eqref{distl-eqn:nn-policy-sampled} using the control objective \eqref{distl-eqn: control-obj} (and regularisation term for $\kappa_{reg}$).  RMSProp \citep{RMSProp} is used for training, and was chosen by benchmarking several choices from Flux.jl \citep{Flux.jl-2018} on the problem.  

Due to the relatively high dimensionality of the system (50 states) and as the training was performed on a laptop (16 GB RAM, i5-101310U processor) to perform the training efficiently we followed the following strategy:

\begin{enumerate}
    \item $\pi_{z_0}$ is estimated from the closed-loop data.\\
    As described in section \ref{distl-sec: disturb-and-noise} we simulate the column under control of the MPC policy to find the typical operating region of the column (see Figure \ref{distil-fig:violin-mpc-column}). Using this data we fit a multi-variate normal distribution to the temperatures of the column and a separate multi-variate normal distribution to the hold-ups of the column\footnote{Separation of the temperatures and hold-ups is conservative, and is to ensure that the controller doesn't somehow learn the initial hold-up from the initial temperature profile.}. As a compact probability distribution is required these are truncated at $\pm 3$ standard deviations from the mean.
    
    \item $\pi_{z_0}$ is discretised to yield $\pi^d_{z_0}$  by taking 1000 quasi-random samples.\\
    We wish to use quasi-random samples and not random samples as they cover the domain more evenly at low number of samples. Generating quasi-random samples with correlations is difficult to do directly. We first use a Sobol sequence to generate points from the marginalised cumulative probability distribution of each state. Correlations between these points are then induced by the Iman-Connover method \citep{iman1982-iman-connover-method}. 

    \item Noise is similarly sampled from the assumed distributions to give $\pi^d_{\eta}$ \\
    Note that as the noise is uncorrelated, correlations are not induced by the Iman-Connover method.
    
    \item RMSProp is used to train the controller. At each iteration $n_s$ points are independently sampled from $\pi^d_{z_0}$ and  $\pi^d_{\eta}$.\\
    For the first 2000 iterations we use $n_s=1$, and then perform 750 iterations with $n_s=2$ (with $w_s=0.5$). The number of iterations is chosen by picking a point on a plot of the objective function versus the number of iterations at which the objective stopped improving. The last 750 iterations serve to reduce the variability of the estimate of $\bar{\theta}^*$. The objective gradient is calculated by  reverse-mode automatic-differentiation on the differential equation solver \citep{innes2018-zygote}. 
\end{enumerate}

Lastly if a regularised objective is used then after 4. the entries of $D$ that are smaller than 0.001 are set to zero. Then the training is repeated without regularisation with $D$ frozen to its new values, and $\theta$ set back to the initial guess. This is because the elastic net regularisation also influences the expressivity of the network, which is an undesired behaviour.

\section{Results}\label{distil-sec: results}
We now compare the policies trained in the previous section, on controlling the column subject to the disturbance sequence shown in Figure \ref{dist-sec: disturbance sequence}. Section \ref{distil-sec: results-noise} compares the nominal performance of the policies with and without measurement noise. In Section \ref{distil-sec: results-mismatch} we compare how the policies handle model mismatch, first with respect to the specified noise distribution, and then with respect to the implemented control to the column. Although the control policies are not directly interpretable, by examining their outputs we aim to provide some interpretations of their behaviour.

\subsection{Comparison of policies with no model-mismatch}\label{distil-sec: results-noise}
In this section the policies are compared based on their performance of controlling the column subject to a test sequence of disturbances, prepared in the same way as when determining the operating region of the column. Note that as before the column starts from initially equimolar composition on each stage. This  ``start-up'' period is outside of the operating region used in the policy optimisation (see Figure \ref{distil-fig:violin-mpc-column}). Thus, the control policies  $\kappa_{all}$, $\kappa_{reg}$, and $\kappa_{sel}$ will perform poorly in this region. The performance of the control policies are summarised in \cref{dist-tab:objective-values}, with further details given in the following.

\begin{table}[tb]
\centering
\caption{Cumulative closed loop objective value from using the trained policies on the test sequence of disturbances excluding the start-up portion. Noise refers to constant measurement noise along a trajectory. Unlike the other policies, $\kappa_{mpc}$ uses full state feedback and hence does not have noise entries.}
\label{dist-tab:objective-values}
\begin{tabular}{l|ccc}
 \hline
     & \multicolumn{3}{c}{Cumulative objective}               \\
     & no noise & with noise & averaged with noise \\ \hline 
$\kappa_{mpc}$ & 0.0076     & -           & -                \\     
$\kappa_{all}$ & 0.0092     & 0.0079       & 0.0096                \\
$\kappa_{sel}$ & 0.0087     & 0.0103       & 0.0094                \\
$\kappa_{reg}$ & 0.0142     & 0.0175       & 0.0157            \\ \hline  
\end{tabular}
\end{table}

\subsubsection{$\kappa_{all}$ and $\kappa_{mpc}$ }

\begin{figure}
    \centering
    \includegraphics[width=\linewidth]{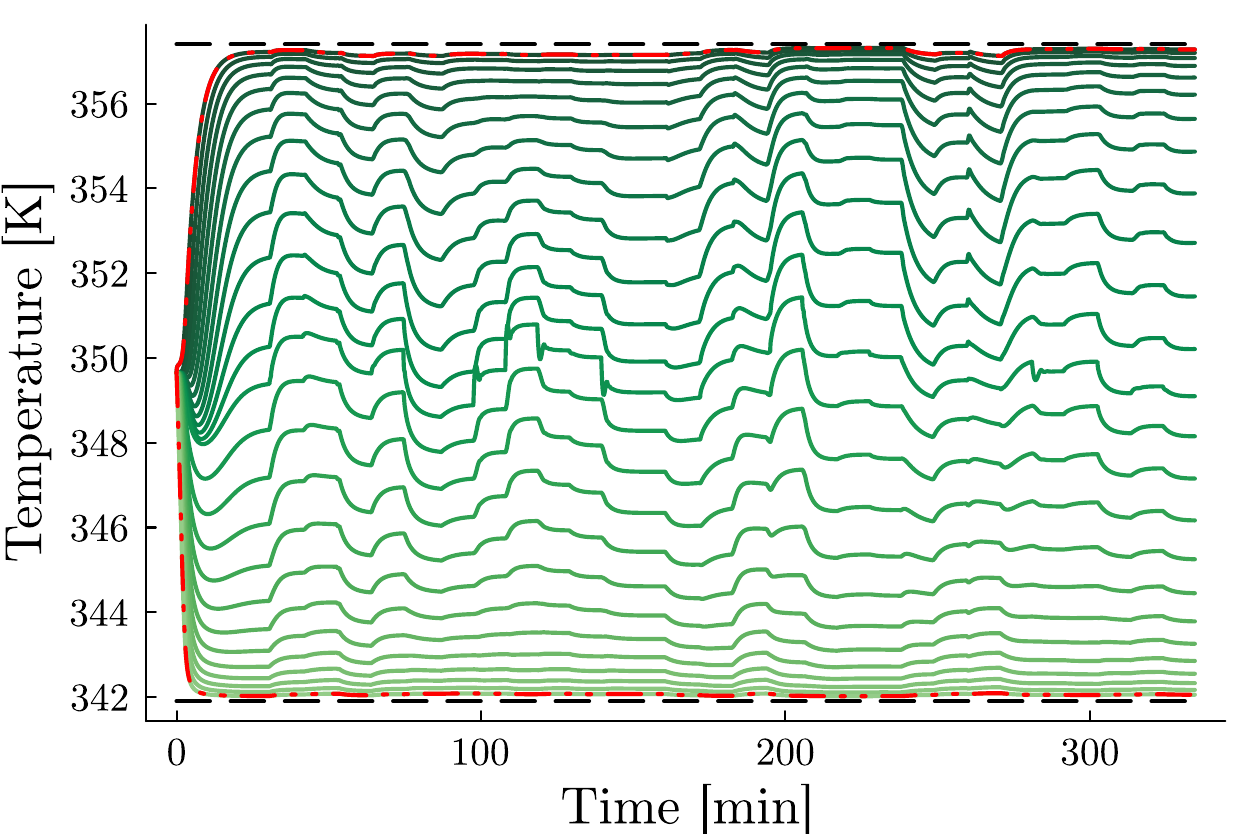}
    \caption{Temperature profile using $\kappa_{all}$. Black dashed lines indicated $T_{bL}$ and $T_{bH}$, red dash-doted lines indicates $T_1$ and $T_{N_T}$, and the green lines indicate column temperatures.}
    \label{distil-fig: Tprof-all-bias-false}
\end{figure}

We first consider the case where  $\kappa_{mpc}$  and $\kappa_{all}$ are used to control the column with no model-mismatch or noise in the measurements, with the resulting temperature profiles shown in Figures \ref{distil-fig: Tprof-mpc-test} and \ref{distil-fig: Tprof-all-bias-false}. By inspection the behaviour of the system is qualitatively similar, although there are differences. For example, between 30-60 minutes $\kappa_{mpc}$ is able to stabilise the system very quickly, while $\kappa_{all}$ has a much larger transient response. 

This is to be expected as with no measurement noise or model error $\kappa_{mpc}$ is able to exactly compensate for (some) incoming disturbances. On the other-hand,  $\kappa_{all}$ does not use the hold-ups of the internal stages of the column (which the MPC does), and thus worse performance should be expected. However the actual difference in the control objective is minor -- the cumulative objective value (excluding the start-up portion) is 0.0076 for  $\kappa_{mpc}$  and 0.0092 for $\kappa_{all}$ (\cref{dist-tab:objective-values}). 
\begin{figure}
     \centering
\begin{subfigure}[t]{0.48\linewidth}
          \centering
    \includegraphics[width=\linewidth]{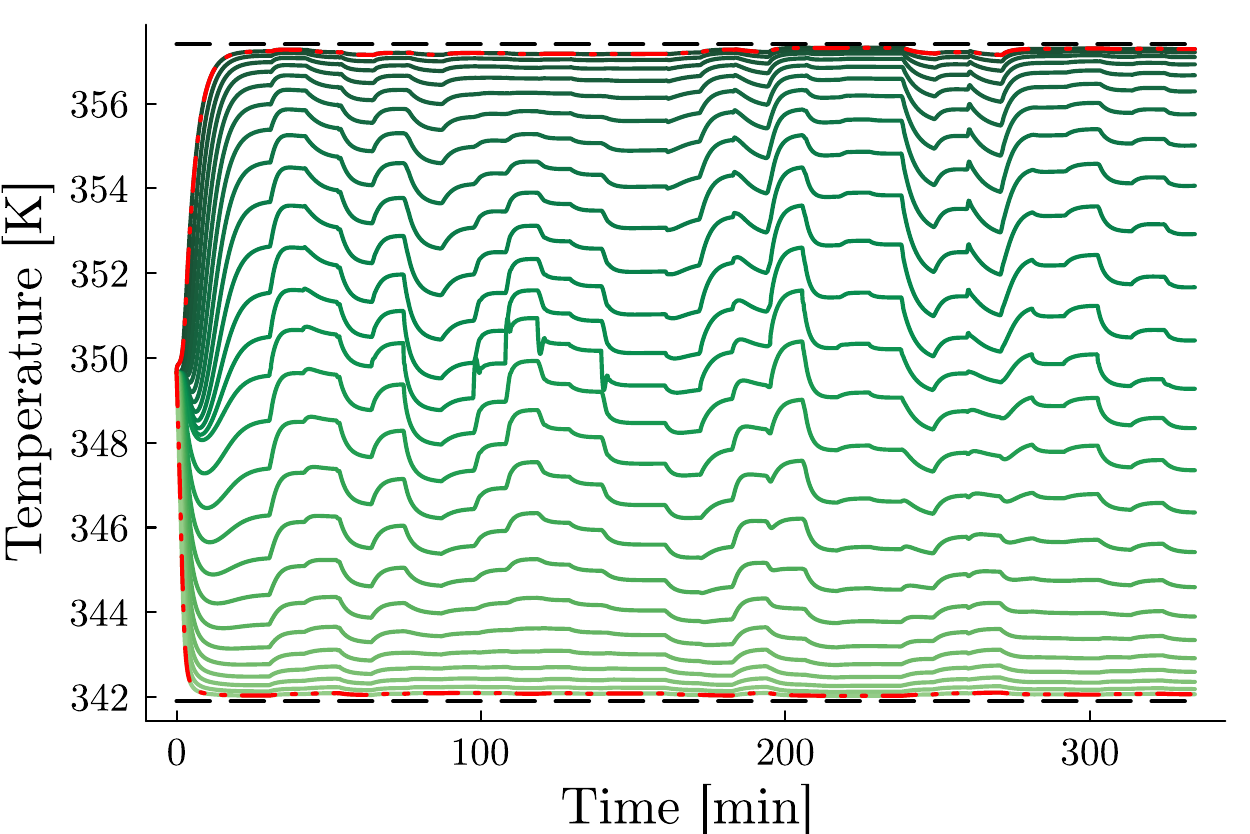}
        \caption{Temperature profile  (constant  bias)}
        \label{distil-fig: Tprof-all-inp-normal-noise}
     \end{subfigure}
     \begin{subfigure}[t]{0.48\linewidth}
          \centering
    \includegraphics[width=\linewidth]{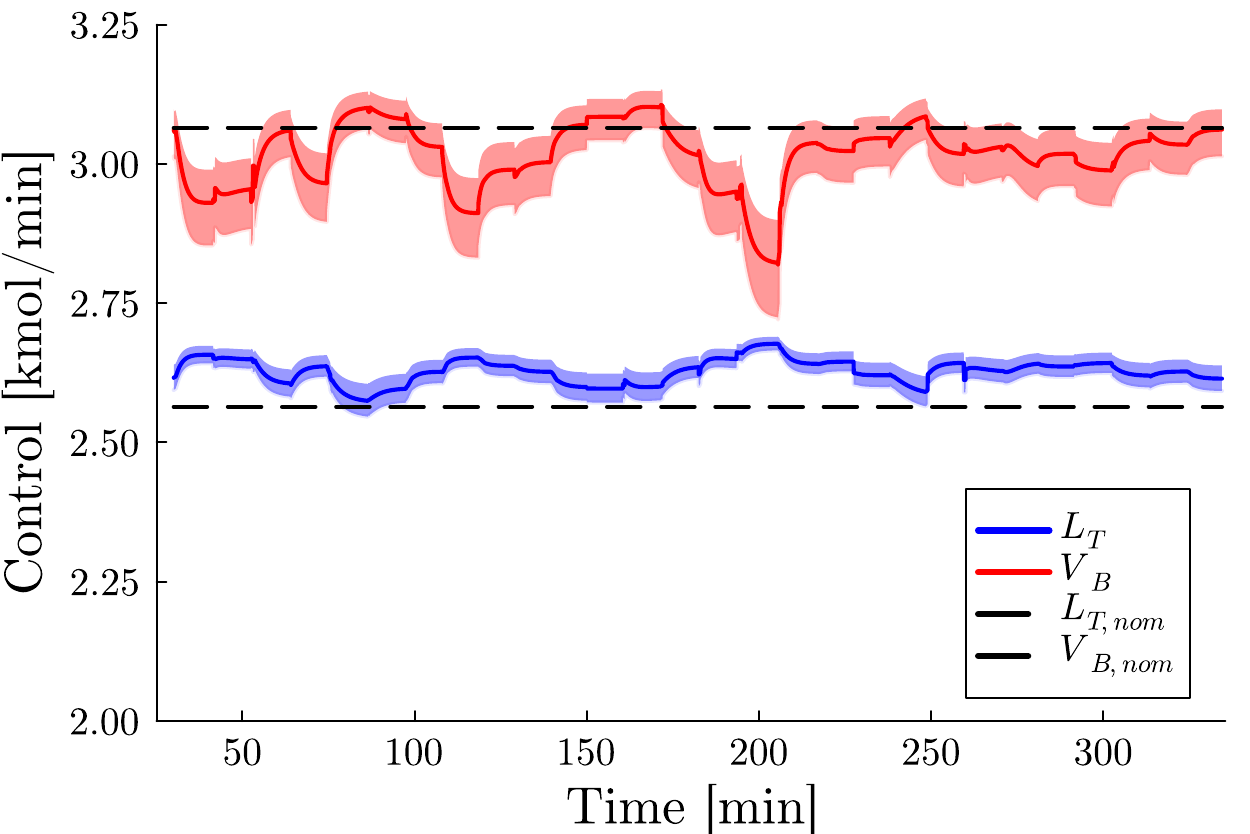}
    \caption{Controller output (varying noise)}
    \label{distil-fig: all-inp-norm-noise}
     \end{subfigure}
     \caption{Closed loop profile of  $\kappa_{all}$  with normally distributed measurement noise. In \cref{distil-fig: Tprof-all-inp-normal-noise} a constant realisation of measurement noise is used across the horizon. In \cref{distil-fig: all-inp-norm-noise} the solid line indicates the nominal control output, while shaded region shows the range of possible control outputs due to normally distributed noise of the measurements.}  
    \label{distil-fig: all-inp-norm-comb}
\end{figure}
To gain further understanding of the behaviour of  $\kappa_{all}$ we can look at the actual control output and how this is influenced by measurement noise, as shown in Figure \ref{distil-fig: all-inp-norm-comb}. An interesting thing to note in that in the control output of $\kappa_{all}$ (\cref{distil-fig: all-inp-norm-noise}) there are sharp spikes at the points of some disturbances -- this means that the controller is directly making use of the feed (disturbance) measurements for feed-forward control. Although this comes with improved performance it does mean that the controller can be more influenced by model error. Despite this, \cref{distil-fig: Tprof-all-inp-normal-noise} shows that in the presence of a constant measurement bias  $\kappa_{all}$ is able to maintain reasonable control of the system. 

\begin{figure}
     \centering
\begin{subfigure}[t]{0.48\linewidth}
          \centering
    \includegraphics[width=\linewidth]{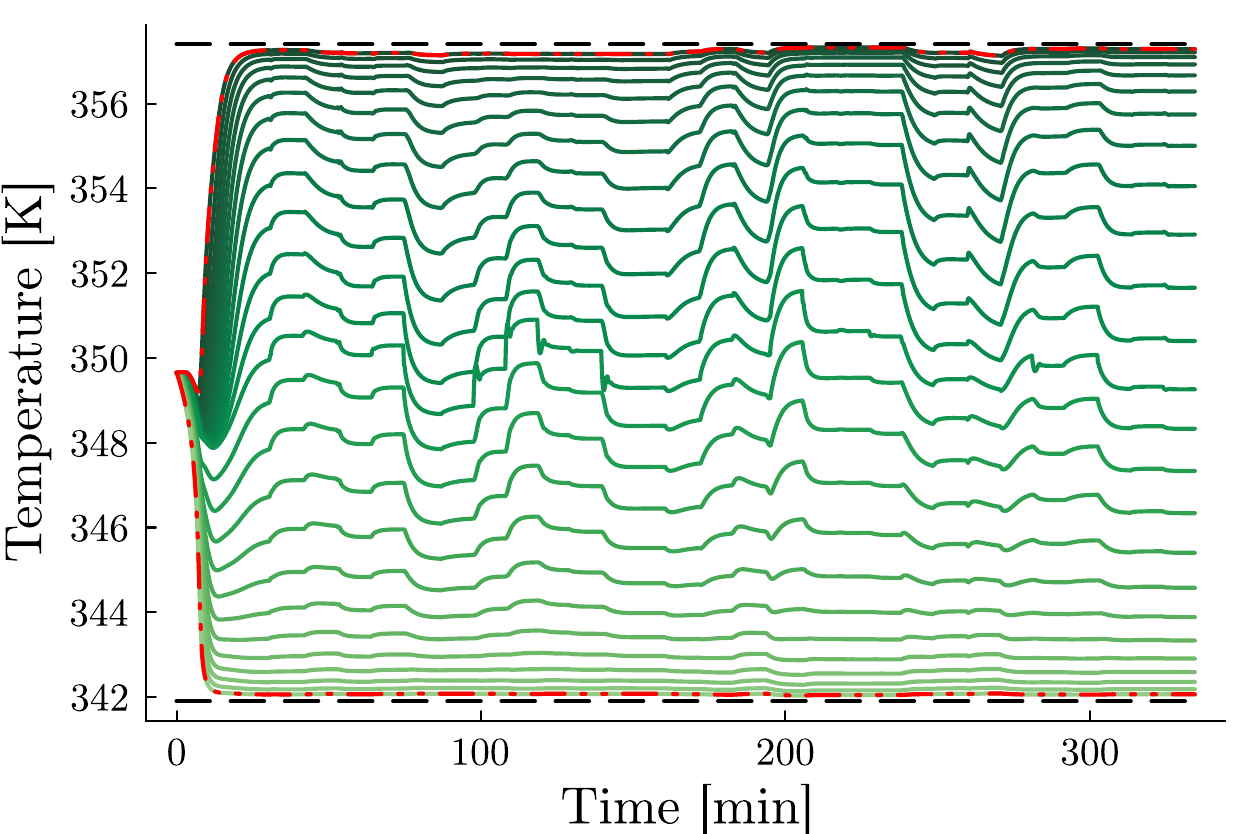}
        \caption{Temperature profile  (constant  bias)}
        \label{distil-fig: Tprof-all-noiseless-inp-normal-noise}
     \end{subfigure}
     \begin{subfigure}[t]{0.48\linewidth}
          \centering
    \includegraphics[width=\linewidth]{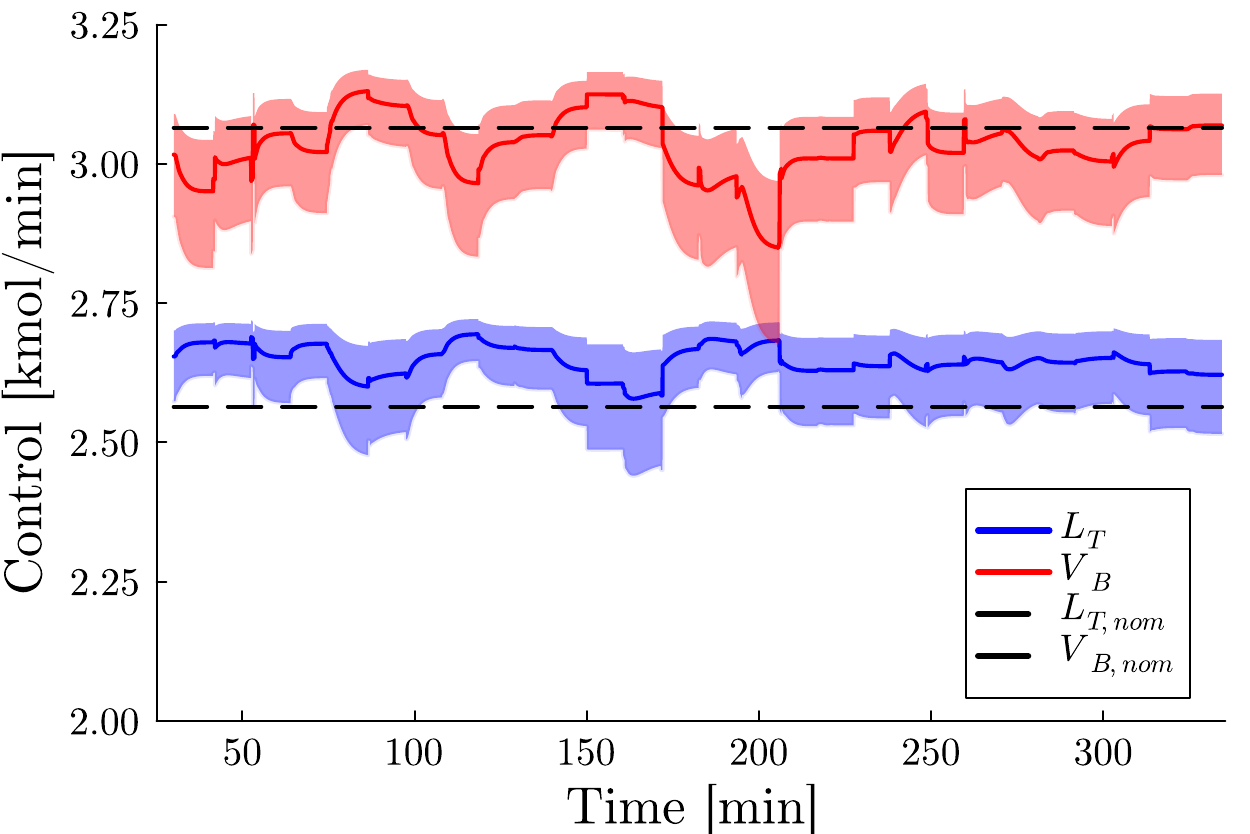}
    \caption{Controller output (varying noise)}
    \label{distil-fig: all-inp-noiseless-norm-noise}
     \end{subfigure}
    \caption{Closed loop profiles using $\kappa^{no\ noise}_{all}$ with normally distributed measurement noise. Green lines indicate column temperatures used by the controller, and grey lines indicate unused column temperatures.} 
    \label{distil-fig: all-inp-noiseless-comb}
\end{figure}

To demonstrate the role of including noise in the training process, we train a controller set-up equivalently to $\kappa_{all}$ but without noise in the training,  $\kappa^{no\ noise}_{all}$. Figure \ref{distil-fig: all-inp-noiseless-comb} shows the closed loop performance of using $\kappa^{no\ noise}_{all}$. This controller output is clearly much more sensitive to noise, and in addition, the spikes in \cref{distil-fig: all-inp-noiseless-norm-noise} are much more pronounced. This difference is expected --   $\kappa^{no\ noise}_{all}$ is more aggressive in its feedforward behaviour at it ``expects'' perfect measurements of the disturbance and column. However, we would also like to note that despite the higher sensitivity of   $\kappa^{no\ noise}_{all}$, the controllers have very similar performance when examining the temperature profiles.

\begin{figure}[htb]
     \centering
     \begin{subfigure}[t]{0.48\linewidth}
          \centering
    \includegraphics[width=\linewidth]{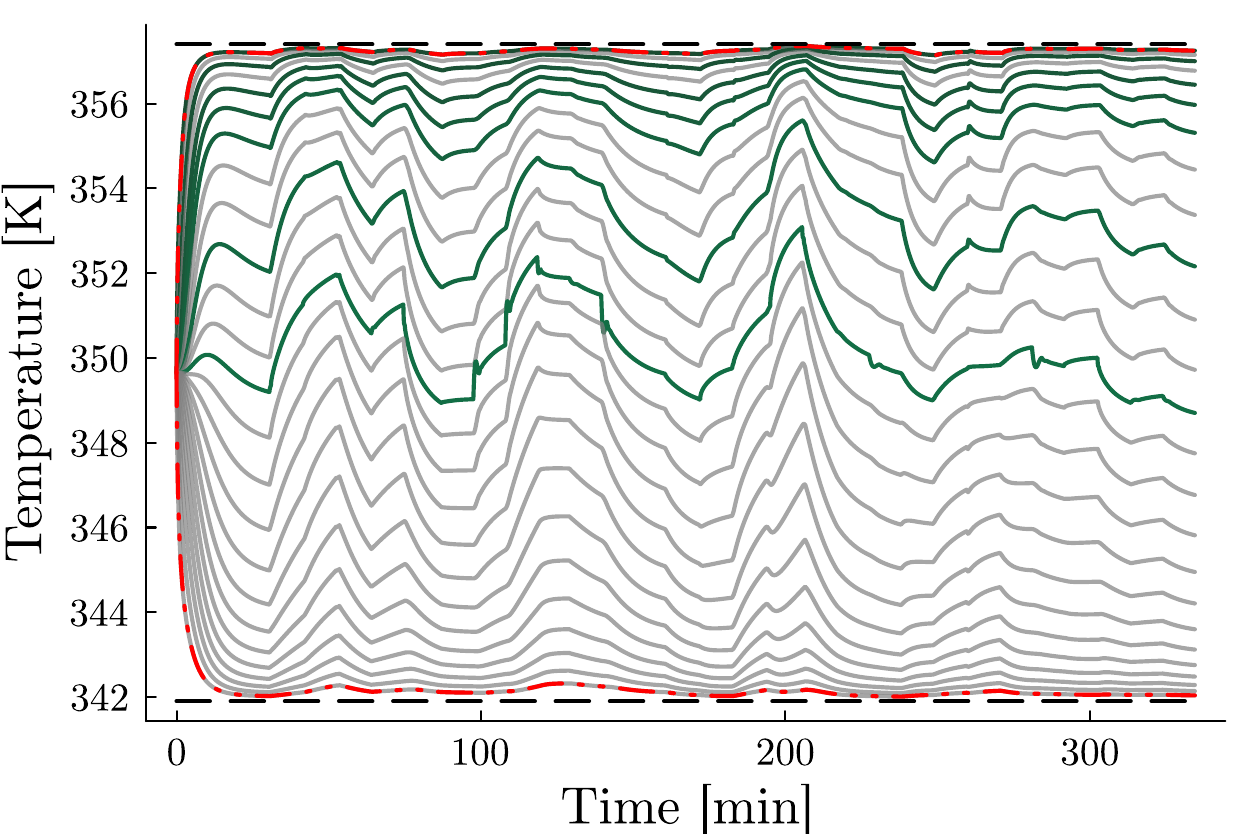}
        \caption{Temperature profile  (constant  bias)}
        \label{distil-fig: Tprof-regul-inp-normal-noise}
     \end{subfigure}
\begin{subfigure}[t]{0.48\linewidth}
          \centering
    \includegraphics[width=\linewidth]{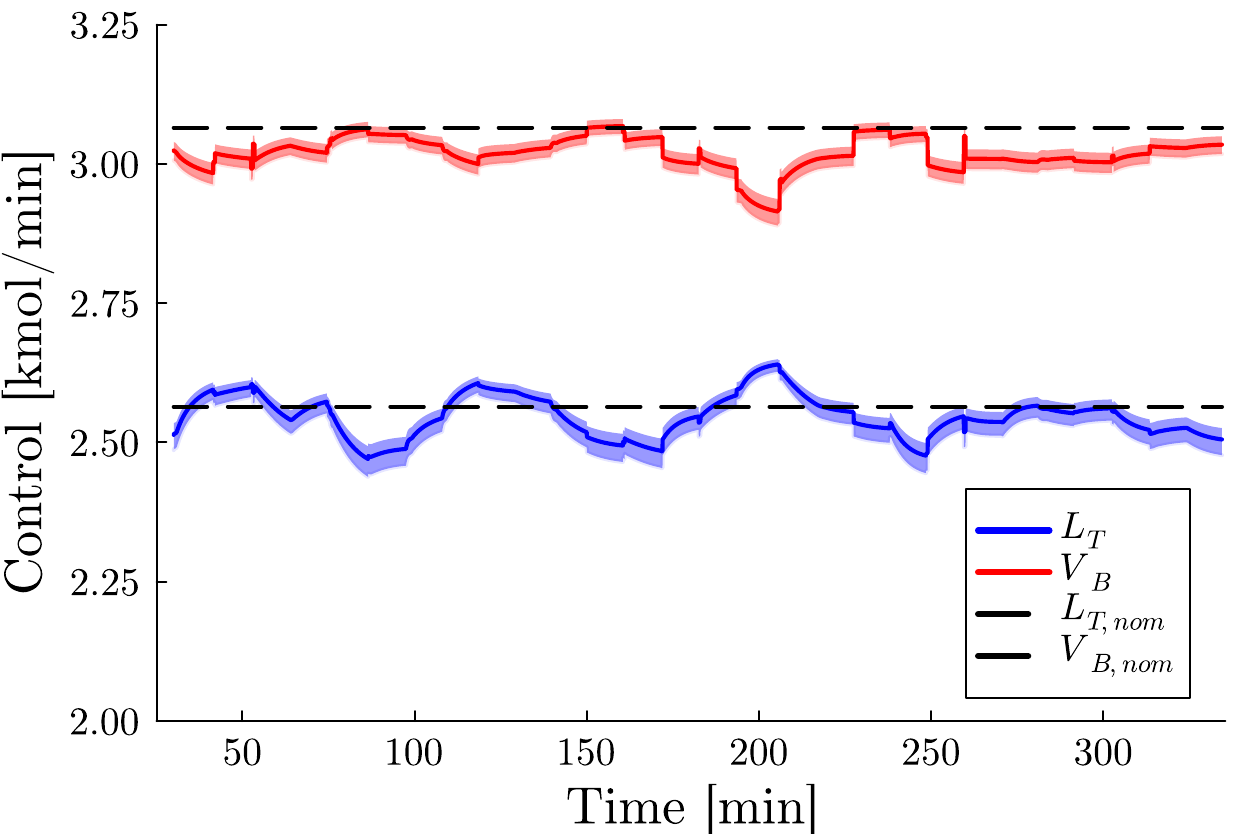}
    \caption{Controller output (varying noise)}
        \label{distil-fig: regul-inp-norm-noise}
     \end{subfigure}
\caption{Closed loop profile of $\kappa_{reg}$ with normally distributed measurement noise.}      \label{distil-fig: regul-inp-norm-comb}
\end{figure}
\subsubsection{Regularised input selection: $\kappa_{reg}$}

$\kappa_{reg}$ is optimised with the elastic net parameters $\lambda_1 = 0.01$, $\lambda_2 = 0.99$.  After the first stage of optimisation, measurement selection is performed by checking for entries of $D$ larger in magnitude than 0.001. These hyper-parameters were chosen as being demonstrative of the kind of measurement selection that can occur, and were not selected based on the closed loop performance. The training yielded nine selected measurements:  $T_{1}$, $T_{3}$,  $T_{5}$, $T_{6}$, $T_{7}$, $T_{10}$, $T_{13}$, $F$, and  $q_F$. Upon first glance this selection may seem strange -- the seven temperature measurements are from the bottom section of the column, and the other two measurements relate to the feed. 

The repeated bottom temperature measurements may allow the controller to combine measurements for less sensitivity to noise, but this doesn't explain why only bottom measurements are used. Note that as the feed is mainly liquid (even with the disturbances to $q_F$), and upon entering the column most of the feed immediately proceeds into the bottom section.  Combined with the liquid flow dynamics this means that multiple temperature measurements in the bottom section of the column provides a short term ``record'' of previous disturbances, and are thus very informative of the overall column state (assuming that there is no model mismatch). By this reasoning one should expect $\kappa_{reg}$ behaviour substantially in a feedforward manner as the disturbances are measured exactly, and indirectly the column state and disturbance history are measured. Considering that the training problem is to reject disturbances while penalising the use of many  measurements, this behaviour is not surprising as feedforward control will yield tighter control than pure feedback control in the absence of model error.

The closed loop simulations of  $\kappa_{reg}$ is shown in Figure  \ref{distil-fig: regul-inp-norm-comb}. The feedforward nature of $\kappa_{reg}$  can be seen by examining the nominal control profile  (Figure  \ref{distil-fig: regul-inp-norm-noise}) which varies very slowly between disturbances, while having sharp changes whenever there is a large disturbance in the feed. In general,  $\kappa_{reg}$ shows much less variation than $\kappa_{all}$ both in the nominal control response, and under the influence of noise.  

The temperature profile has relatively sharp and rapid changes compared to  compared to Figure \ref{distil-fig: all-inp-norm-comb}. Visually in some places $\kappa_{reg}$ appears to perform better than $\kappa_{all}$, however the cumulative control objective over the test horizon using $\kappa_{reg}$ is worse than using $\kappa_{all}$ (see \cref{dist-tab:objective-values}).  

\subsubsection{Control policy with manual input selection: $\kappa_{sel}$}
\begin{figure}
     \centering
     \begin{subfigure}[t]{0.48\linewidth}
          \centering
    \includegraphics[width=\linewidth]{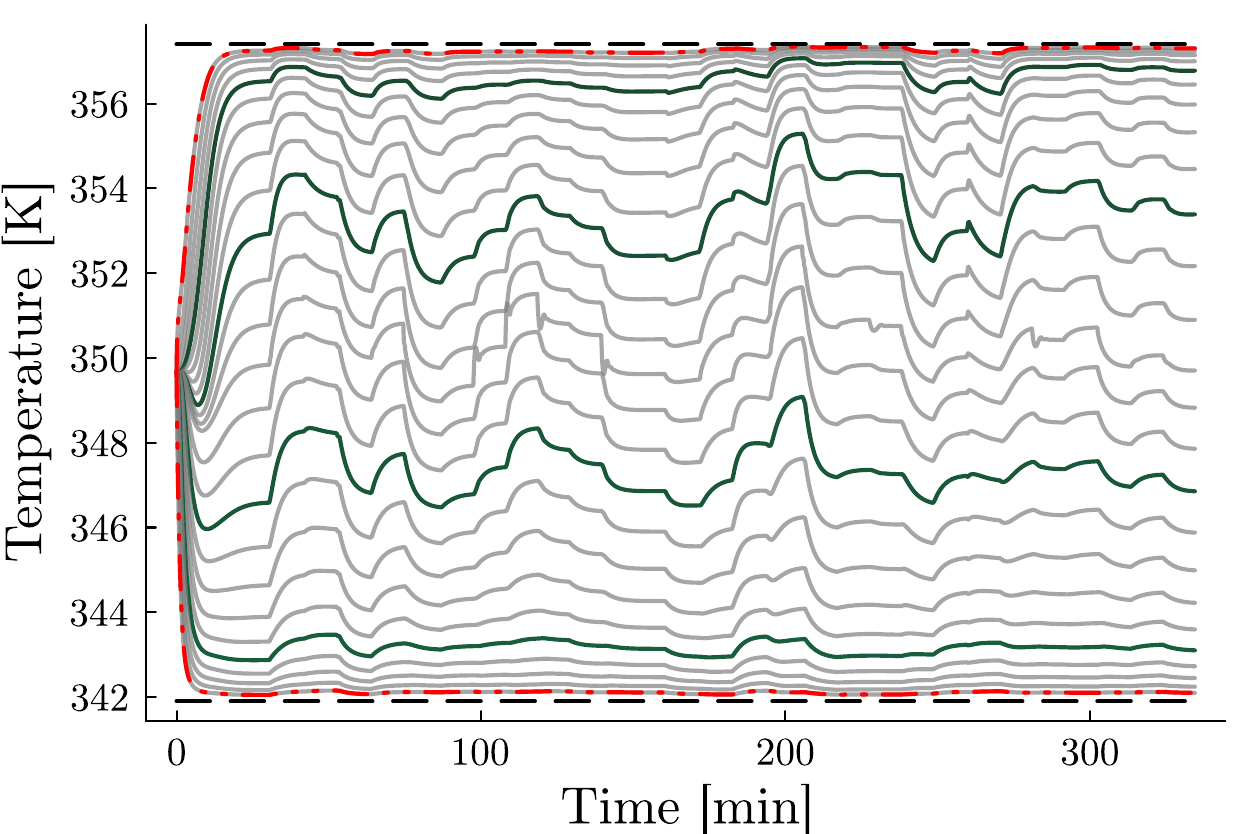}
        \caption{Temperature profile  (constant  bias)}
        \label{distil-fig: Tprof-sel-inp-normal-noise}
\end{subfigure}
\begin{subfigure}[t]{0.48\linewidth}
          \centering
    \includegraphics[width=\linewidth]{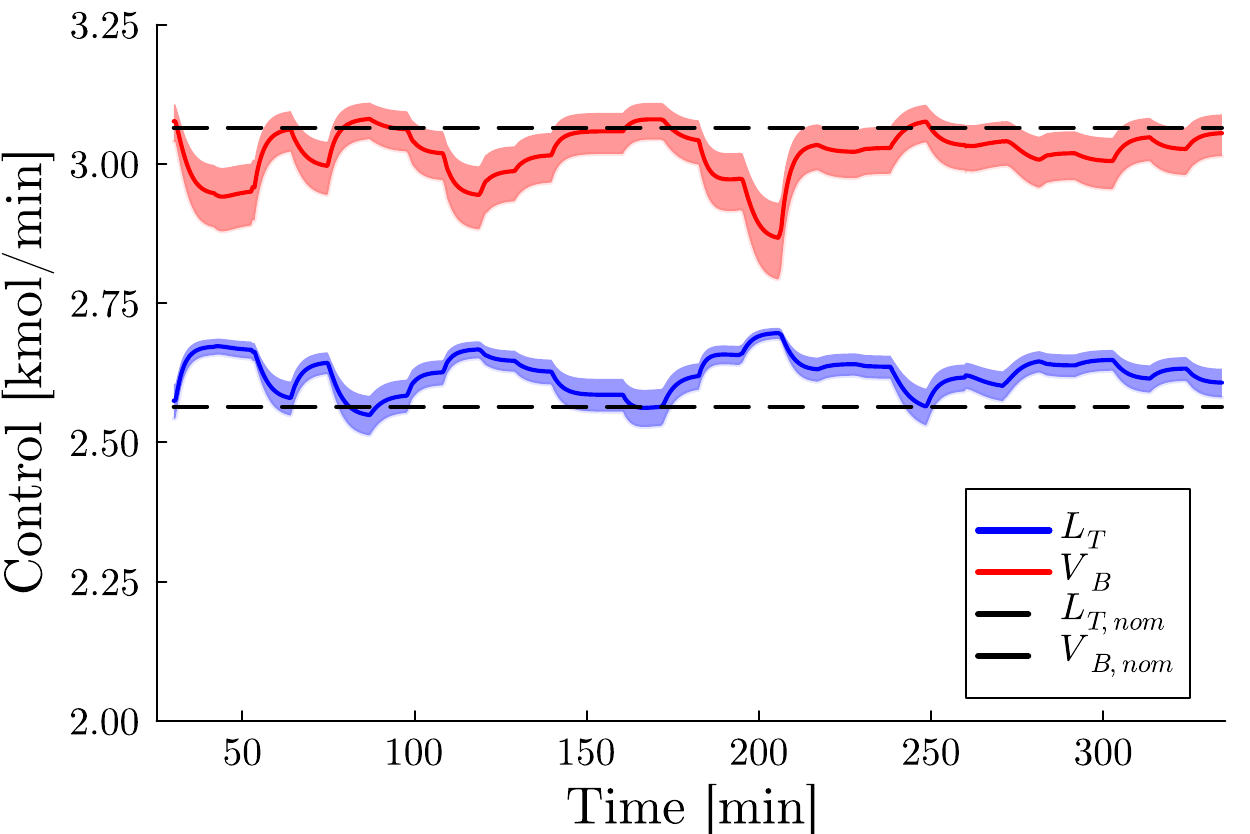}
    \caption{Controller output (varying  noise).}
        \label{distil-fig: sel-inp-norm-noise}
     \end{subfigure}
\caption{Closed loop profile of $\kappa_{sel}$ with normally distributed measurement noise.}    \label{distil-fig: sel-inp-norm-comb}
\end{figure}

Unlike $\kappa_{reg}$ where we regularise for input selection, for   $\kappa_{sel}$  the inputs are chosen as $\zeta = [\bar{T}_5, \bar{T}_{10}, \bar{T}_{16}, \bar{T}_{21}]$.
As discussed earlier, the motivation behind this highly reduced set of temperature measurements is that these are similar temperatures that may be used in classic distillation control strategies. In addition note that this removes the possibility of the controller performing feedforward control on the disturbances.  The resulting temperature and control profiles using $\kappa_{sel}$  are shown in Figure \ref{distil-fig: sel-inp-norm-comb}. 

In general there is more variation in the control output than $\kappa_{reg}$ , but less than $\kappa_{all}$. However, the temperature profiles are much less sensitivity to noise than $\kappa_{reg}$. Interestingly, the temperature and control profile of $\kappa_{sel}$ and $\kappa_{all}$ are very similar. In addition, the cost of using these controllers is very similar (\cref{dist-tab:objective-values}). In particular,  $\kappa_{sel}$ seems to be giving a smoother version of $\kappa_{all}$, as it lacks the initial kick from the feed measurements provided to $\kappa_{all}$.  What this suggests is that $\kappa_{sel}$ contains the key elements of the ``feedback part" of $\kappa_{all}$ while neglecting the feedforward part on the disturbances. Due to the time scale, the slightly slower response of $\kappa_{sel}$ compared to $\kappa_{all}$ does not make a significant difference to the cumulative objective.

\subsection{Controller sensitivity to  mismatch}\label{distil-sec: results-mismatch}

It is well known that optimal control policies can potentially be sensitive to mismatch between the model used in their solution and the true system. %
In this section we compare the controllers when: (1) the measurement noise is mispecified and (2) when there is a multiplicative output disturbance. We find that of the trained controllers $\kappa_{sel}$ is much more robust to these disturbances and is better able to regulate the system. However, as all control schemes do not have an integrator, offset free control is not achieved.

\subsubsection{Mispecification of noise distribution}
\begin{figure}[p]
     \centering
     \begin{subfigure}[t]{0.48\linewidth}
          \centering
    \includegraphics[width=\linewidth]{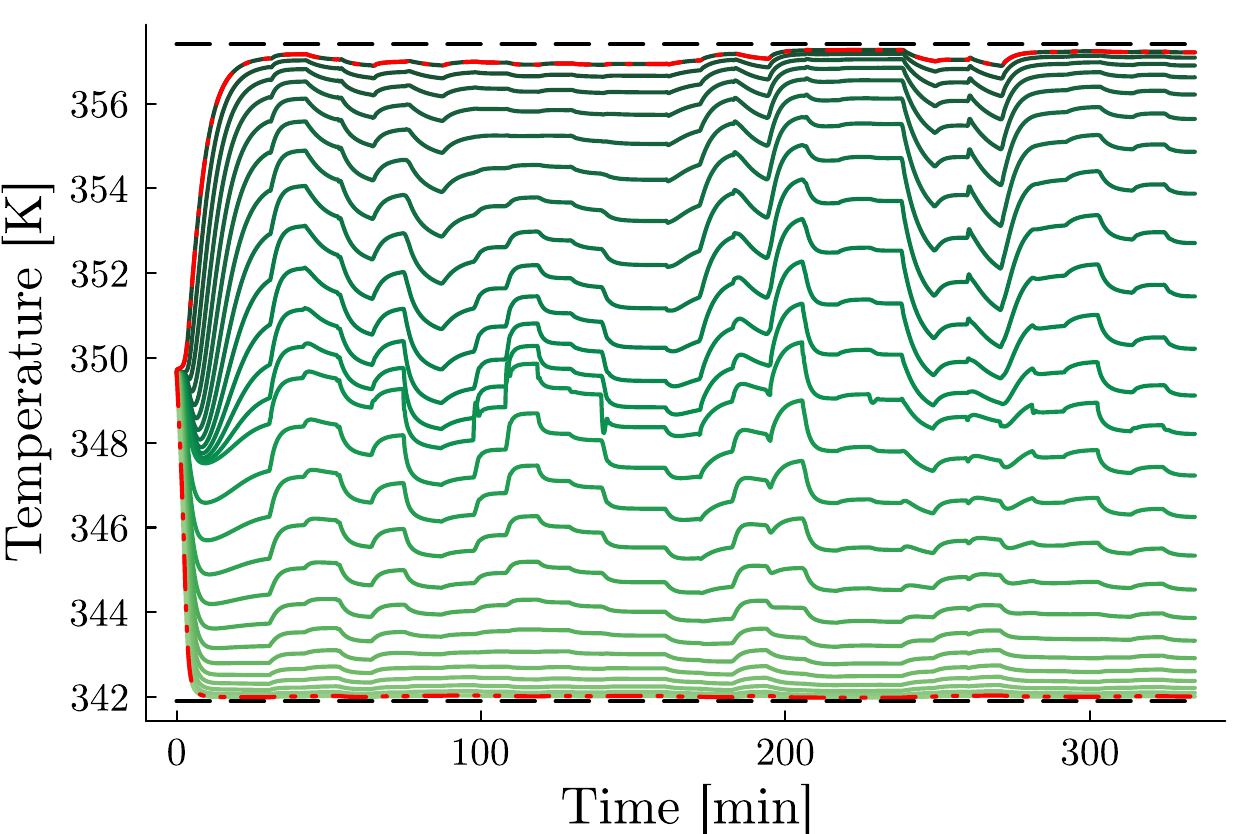}
        \caption{Temperature profile  (constant  bias)}
    \label{distil-fig: Tprof-all-inp-worst-noise}
     \end{subfigure}
    \begin{subfigure}[t]{0.48\linewidth}
          \centering
    \includegraphics[width=\linewidth]{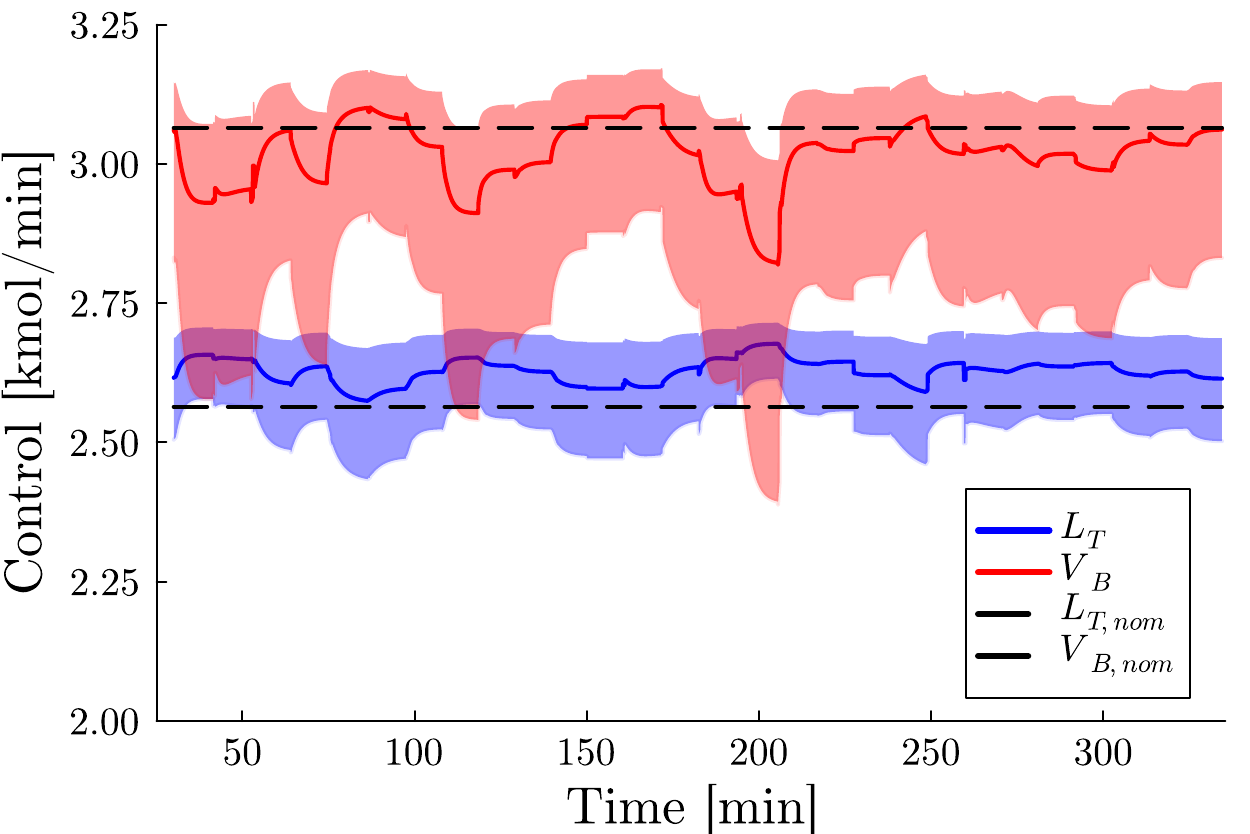}
    \caption{Controller output (varying noise)}
    \label{distil-fig: all-inp-worst-noise}
     \end{subfigure}
    \caption{Closed loop profile of $\kappa_{all}$ with extreme noise realisations.}
    \label{distil-fig: all-inp-worst-comb}
\end{figure}
\begin{figure}[p]
     \centering
     \begin{subfigure}[t]{0.48\linewidth}
          \centering
      \includegraphics[width=\linewidth]{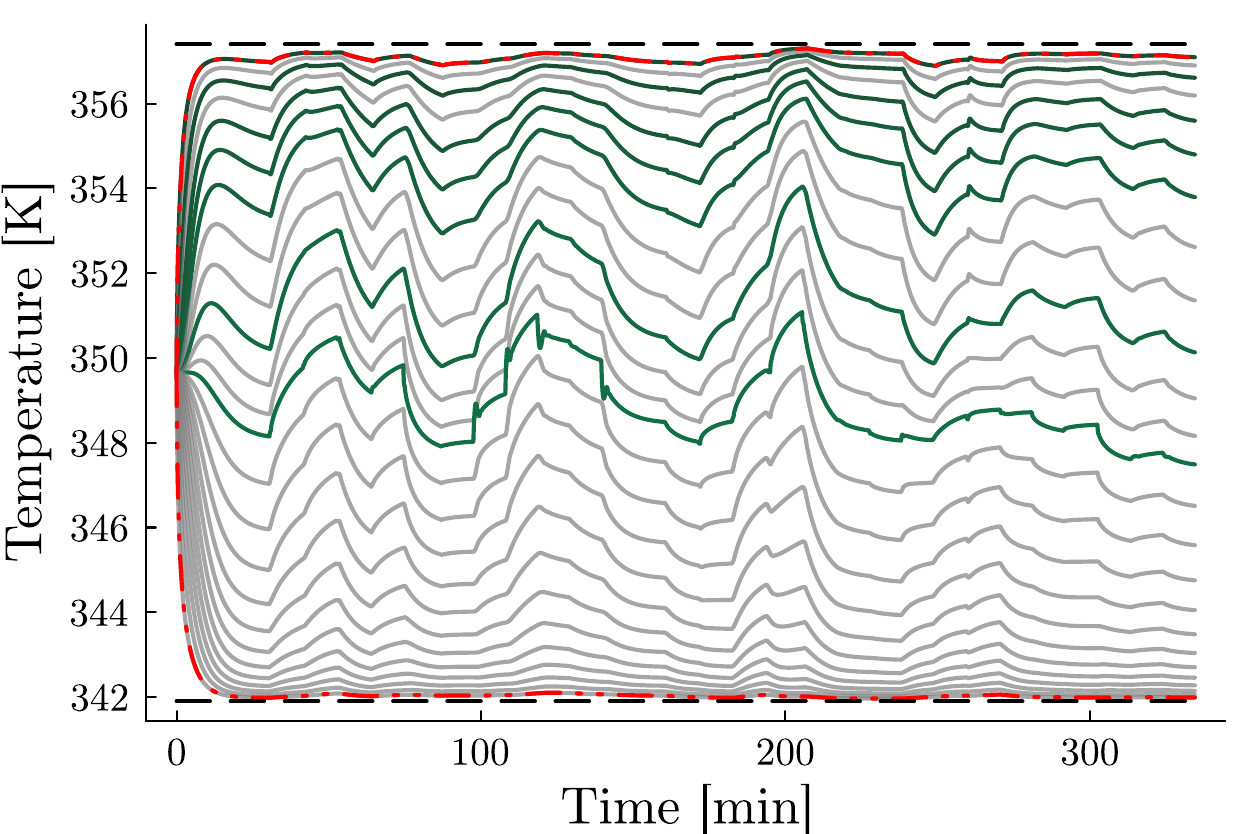}
        \caption{Temperature profile  (constant  bias)}
    \label{distil-fig: Tprof-regul-inp-worst-noise}
     \end{subfigure}
          \begin{subfigure}[t]{0.48\linewidth}
          \centering
    \includegraphics[width=\linewidth]{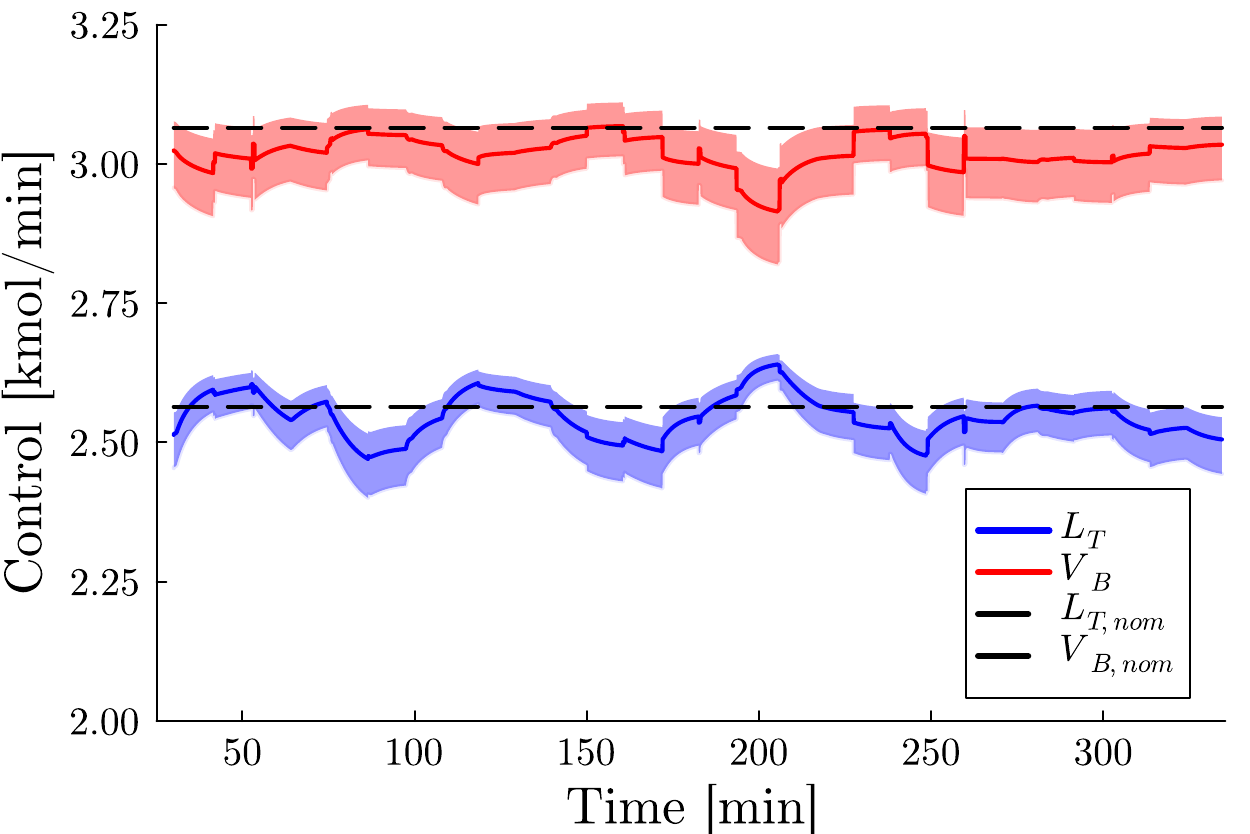}
    \caption{Controller output (varying noise)}
    \label{distil-fig: regul-inp-worst-noise}
     \end{subfigure}
\caption{Closed loop profile of $\kappa_{reg}$ with extreme noise realisations.}    \label{distil-fig: regul-inp-worst-comb}
\end{figure}
\begin{figure}[p]
     \centering

     \begin{subfigure}[t]{0.48\linewidth}
          \centering
      \includegraphics[width=\linewidth]{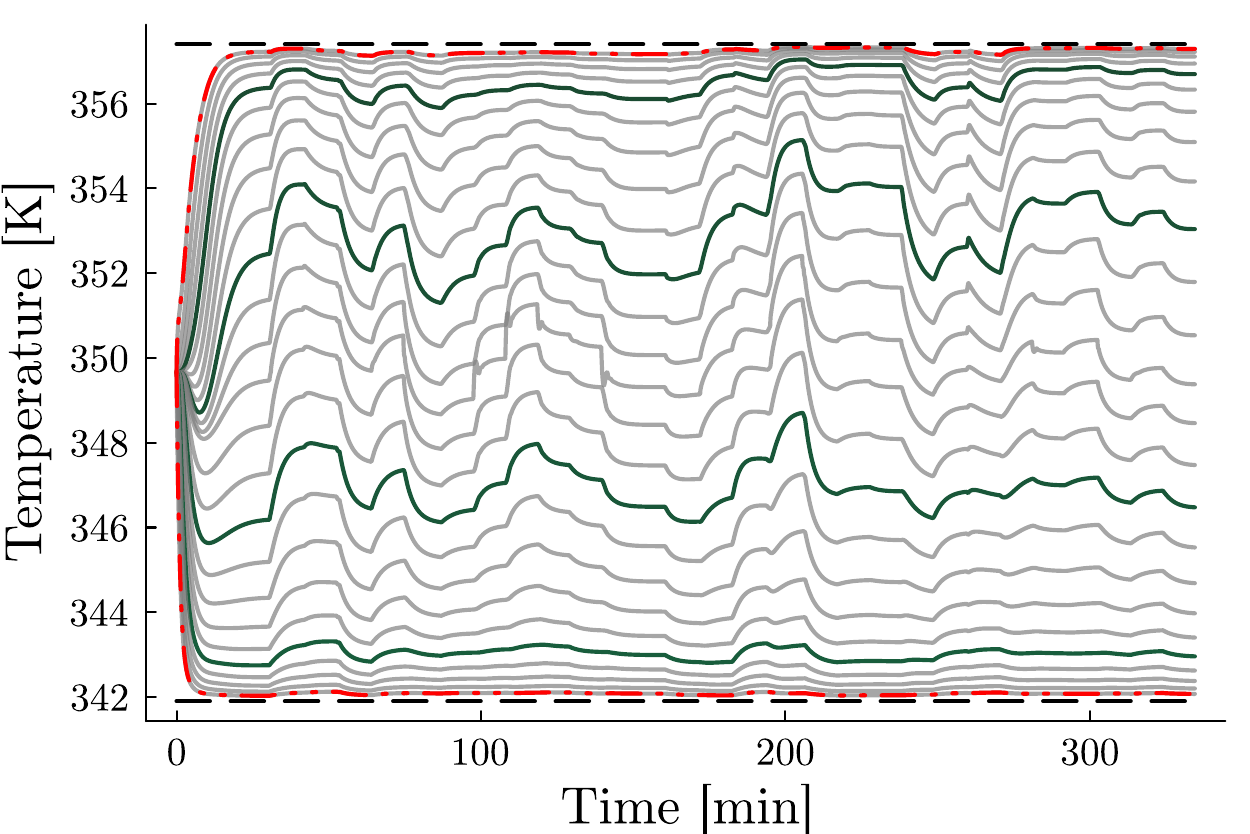}
        \caption{Temperature profile  (constant  bias)}
    \label{distil-fig: Tprof-sel-inp-worst-noise}
     \end{subfigure}
          \begin{subfigure}[t]{0.48\linewidth}
          \centering
    \includegraphics[width=\linewidth]{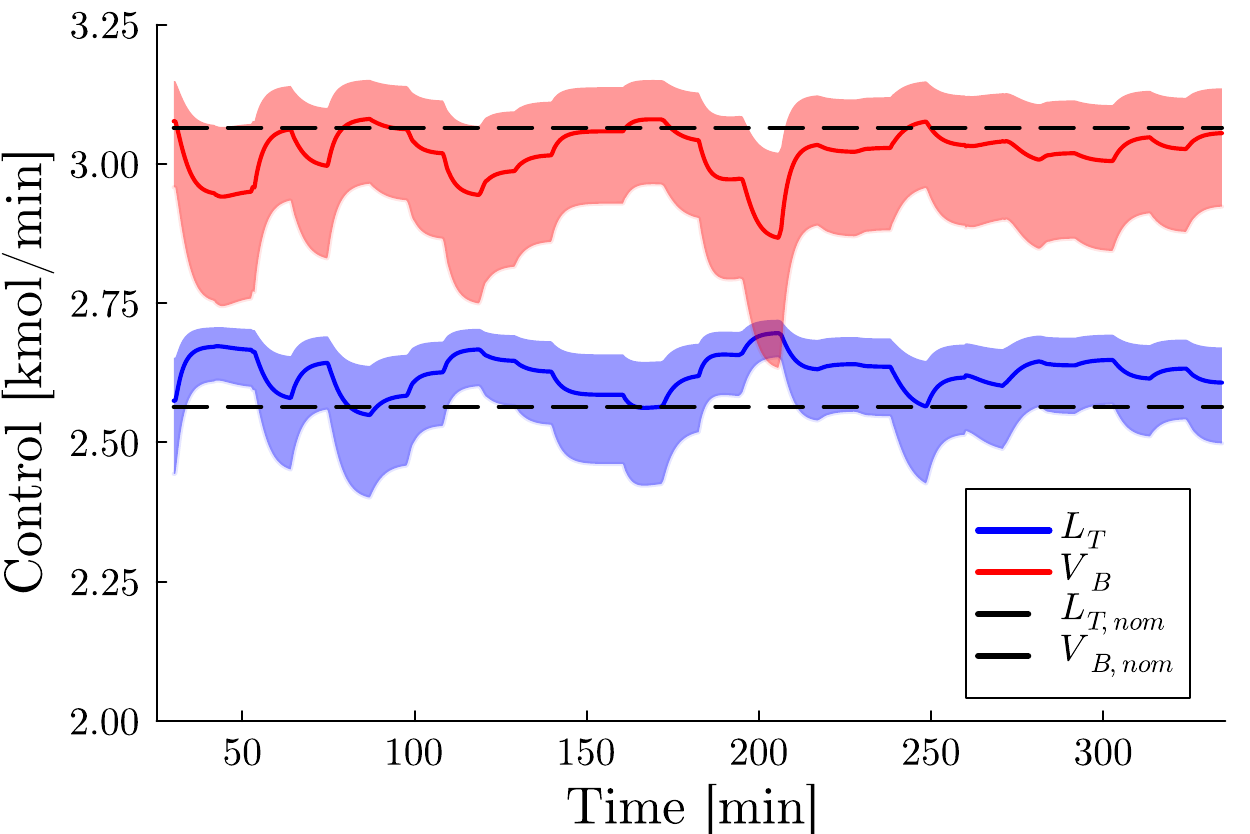}
    \caption{Controller output (varying noise)}
    \label{distil-fig: sel-inp-worst-noise}
     \end{subfigure}
\caption{Closed loop profile of $\kappa_{sel}$ with extreme noise realisations.}    \label{distil-fig: sel-inp--worst-comb}
\end{figure}

 \cref{distil-fig: all-inp-worst-comb,distil-fig: sel-inp--worst-comb,distil-fig: regul-inp-worst-comb} show the closed loop response of the system when instead of normally distributed noise, the noise realisations are exclusively at the extremes of the truncated probability distributions. This is an incredibly unlikely possibility with respect the normally distributed noise and serves as an example of severe mispecification of the noise characteristics of the system.
 
 $\kappa_{all}$,  \cref{distil-fig: all-inp-worst-comb}, has the most variable controller output due to noise and also has the worst temperature regulation. A potential explanation is that as $\kappa_{all}$ has access to many measurements $\kappa_{all}$ may compensate for the noise by combining measurements together.  When the assumption behind the noise is false then this compensation does not work well and results in a large variance in the controller output, as the controller did not learn to be less sensitive to variations. Additional support for this is noting that $\kappa_{reg}$  and $\kappa_{sel}$ are much less sensitive to noise when examining both the controller output and temperature profiles.

Comparing $\kappa_{reg}$  and $\kappa_{sel}$ it is interesting to note that despite $\kappa_{reg}$ showing more variation with noise than $\kappa_{sel}$  (\cref{distil-fig: sel-inp-norm-noise,distil-fig: regul-inp-worst-noise}), in terms of regulation $\kappa_{sel}$ performs better. As noted in the previous section this is likely because $\kappa_{reg}$ primarily makes changes based on the feed measurements, and only small changes based on the temperatures.

\subsubsection{Multiplicative model error}

\begin{figure}[tb]
    \centering
    \begin{subfigure}[t]{0.49\linewidth}
    \centering
    \includegraphics[width=\linewidth]{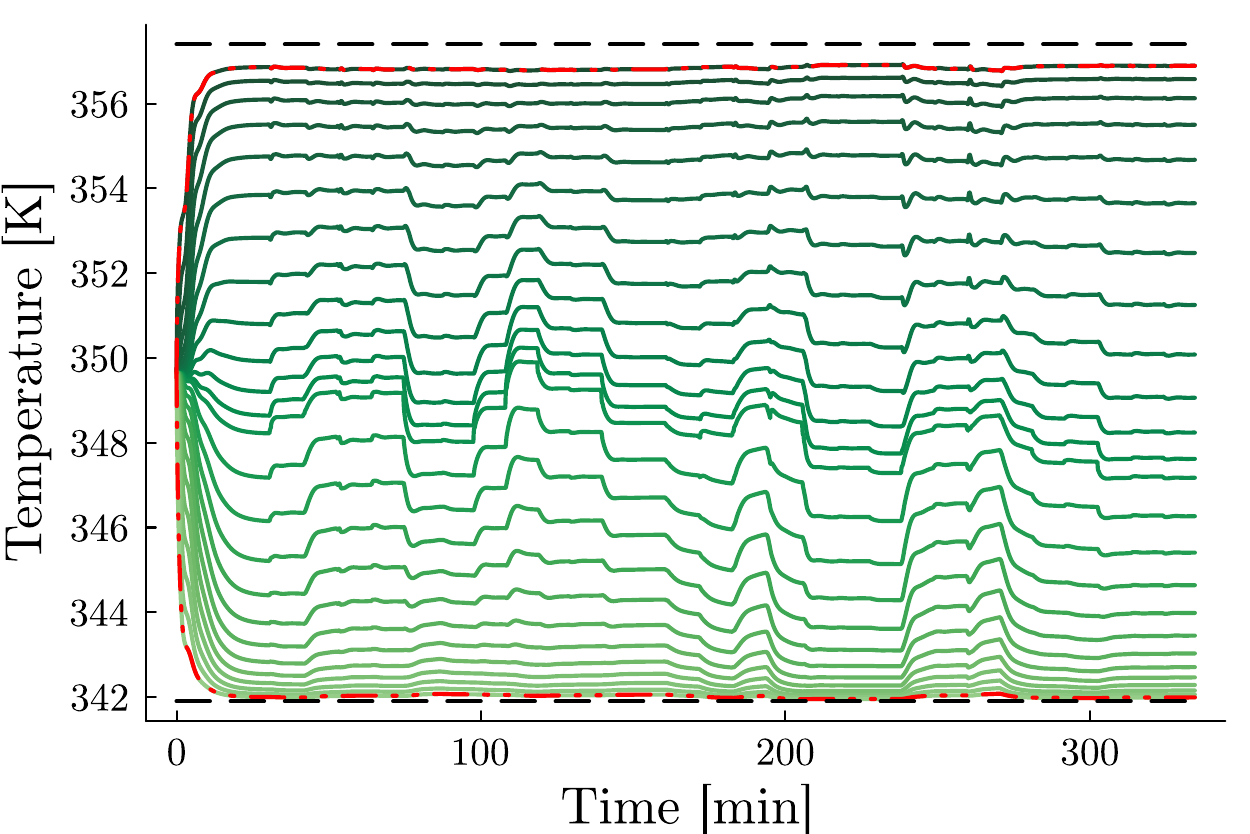}
      \caption{$\kappa_{mpc}$}
    \label{distil-fig: gain-err-mpc}
    \end{subfigure}
\begin{subfigure}[t]{0.49\linewidth}
    \centering
    \includegraphics[width=\linewidth]{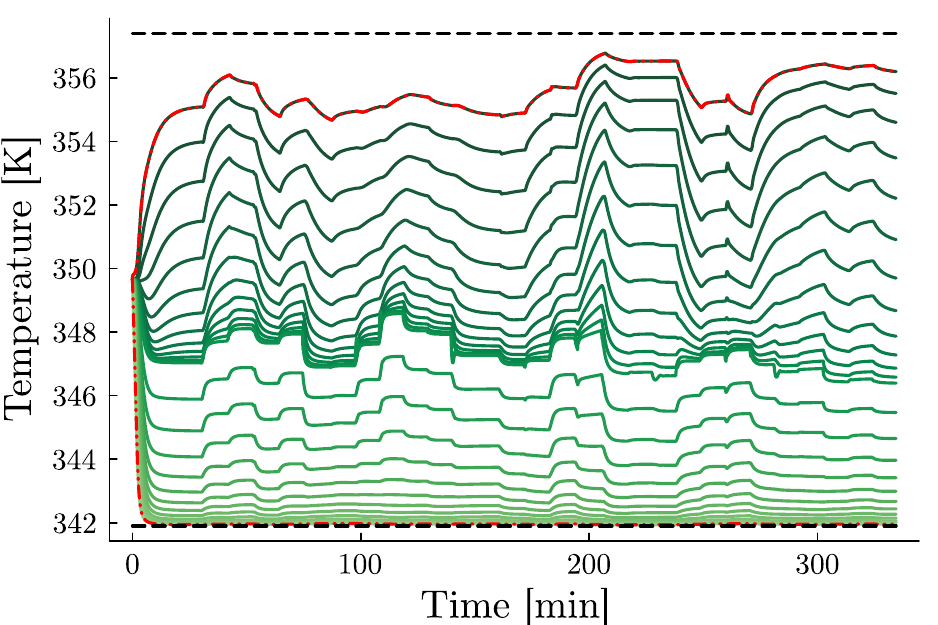}
    \caption{$\kappa_{all}$}
    \label{distil-fig: gain-err-policy-all}
    \end{subfigure}
\begin{subfigure}[t]{0.49\linewidth}
    \centering
    \includegraphics[width=\linewidth]{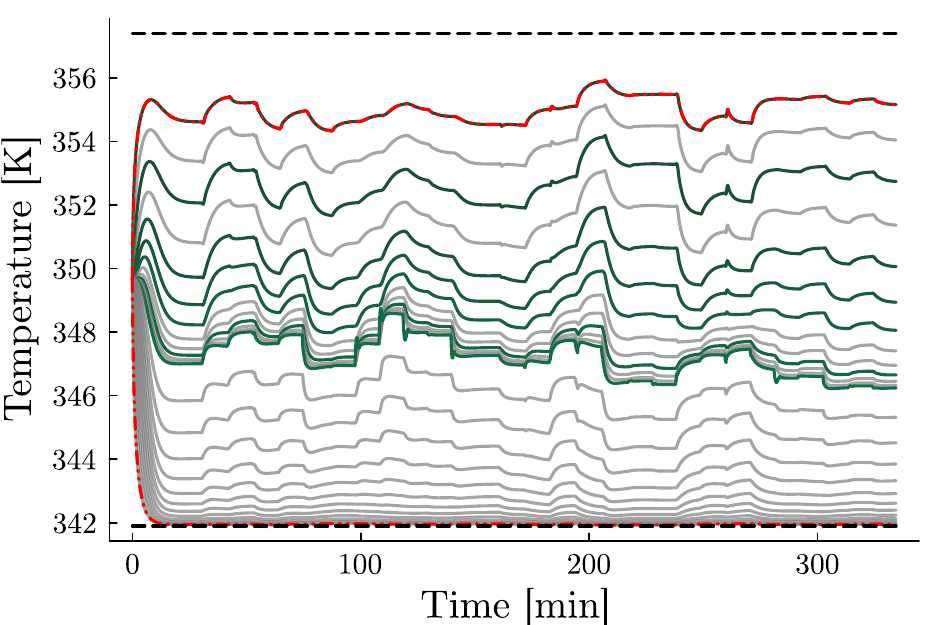}
    \caption{$\kappa_{reg}$}
    \label{distil-fig: gain-err-policy-reg}
    \end{subfigure}
    \begin{subfigure}[t]{0.49\linewidth}
    \centering
    \includegraphics[width=\linewidth]{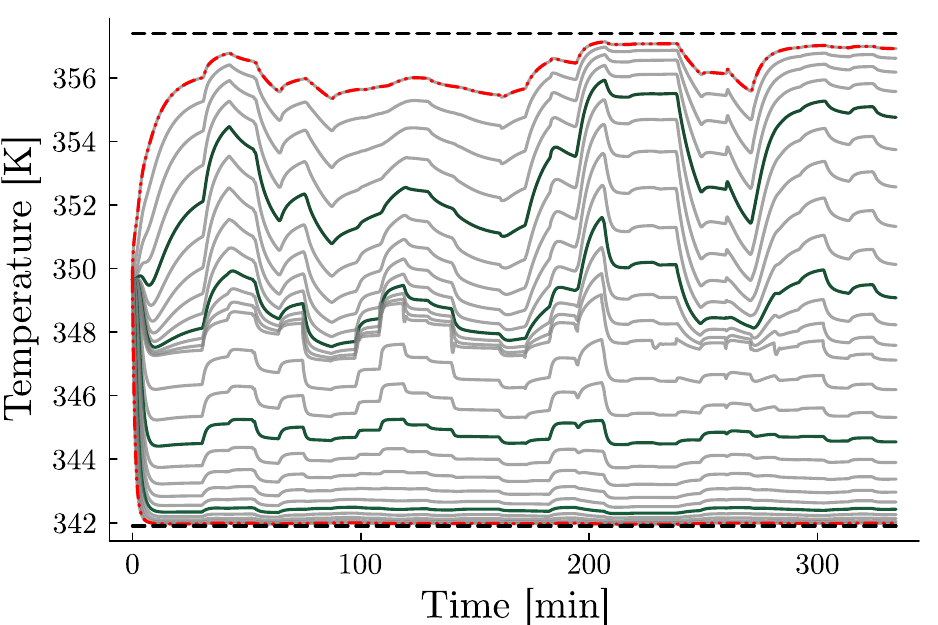}
    \caption{$\kappa_{sel}$}
    \label{distil-fig: gain-err-policy-sel}
    \end{subfigure}
    \caption{Closed loop temperature profiles of the system, with a constant multiplicative output disturbance.}
    \label{distil-fig: gain-err}
\end{figure}

We consider the same model, but assume that due to model-mismatch the reflux and boil up are respectively 10\% more and less than what the policy specifies, i.e.
\begin{equation}
    [L_T,\ V_B
]_{actual} =[1.1L_T,\ 0.9V_B]
\end{equation}
This is a considerably large error, and as the control policies considered in this work are static and not integrating we can reasonably expect significant steady state offset. However, offset free control is not the primary goal, instead we can see how robust the policies are at managing this large, unmodelled disturbance.

The closed loop response of the system, without any measurement noise, and using policies $\kappa_{mpc},\ \kappa_{all},\ \kappa_{reg},$ and $\kappa_{sel}$ is shown in \cref{distil-fig: gain-err}.  Immediately it is clear that the best performing controller is $\kappa_{mpc}$ followed by $\kappa_{sel}$. 
The good performance of $\kappa_{sel}$ is unsurprising given the previous results, as the controller is forced to perform feedback control and robustness is exacerbated by feedforward control.

\section{Conclusion}\label{distil-sec: discussion}\label{distil-sec: conclusion}

We have proposed and demonstrated the closed loop training of (static) measurement based control policies on a distillation case study. This case study is significantly larger compared to prior literature. The key features of the proposed approach are: (1) the controller is trained in closed loop using an optimise-and-learn approach, (2) the controller is trained in an operationally relevant region of the state-space to reduce the computational demands of the training, and (3) using this approach the controller can be trained to use a selection of measurements. Three controllers are trained using different selections of measurements, and their performance is compared for the nominal system (with and without measurement noise) and also for a system with model mismatch. A controller using only 4 measurements along the column is able to perform well, achieving close to MPC performance on the nominal system while being more robust than the other controllers in the example with significant model mismatch.

\subsubsection{Further work}
There are key aspects of the approach that need to be further developed are the selection of measurements for the controller and the incorporation of integration in the control policy.

We have demonstrated that one can set up a training problem to automatically select important measurements for use by a control policy. Through numerical simulation we show that such a control policy ($\kappa_{reg}$) achieves good \textit{nominal} performance. However, this example also shows that doing so can lead to fragile policies that perform poorly beyond the nominal system (\cref{distil-fig: gain-err}). This is because regularisation for measurement selection removes measurements that are not necessary for control of the (assumed perfect) model, which can result in a dramatically worse performance when there is model mismatch. A potential approach to fix this problem is to incorporate uncertainty into the training. However, doing so requires in a meaningful manner requires an accurate description of the uncertainty, and can also dramatically increase the computational complexity of the training.  

Another choice is to use engineering judgement to select measurements that can be used to find good control actions  ($\kappa_{sel}$). Despite using considerably less measurements, this controller was significantly more robust to the others in the numerical examples. Compared to classical control techniques that pair single measurement to control variables, $\kappa_{sel}$ can be trained with multi-variable pairings without specifying the structure of this relationship. However, the central issue of selecting which measurements remains \citep{foss1973critique}.

Without integration, a static controller cannot achieve offset free control by itself. A potential option is to directly include the potential for integration in the controller by allowing it to use an additional state as ``memory''. The challenge with this approach is for the controller to not over-fit during training, and to generalise beyond the model-mismatch, measurement bias, and similar used in the training. Alternatively, integration may be performed separately, e.g. by a disturbance model. In this approach the controller would be trained with the augmented model, and online the disturbance parameters would be provided to the controller to achieve offset free control.
\bibliographystyle{elsarticle-num-names} 
\bibliography{bib}

\end{document}